\documentclass[12pt,reqno,a4paper,fleqn]{amsart}
\usepackage[dvips]{color}
\usepackage{amsmath}
\usepackage{amsxtra}
\usepackage{amscd}
\usepackage{amsthm}
\usepackage{amsfonts}
\usepackage{amssymb}
\usepackage{nccmath}
\usepackage{eucal}
\usepackage{epsfig}
\usepackage{graphics}
\usepackage{dirtytalk}
\usepackage{ulem}
\usepackage[breakable]{tcolorbox}
\usepackage{ifthen}
\usepackage{varioref}

\usepackage{bbold}

\usepackage{enumerate}
\usepackage{hhline}
\usepackage{multirow}

\usepackage[nosort]{cite}
\usepackage{xspace}
\usepackage{setspace}

\usepackage{titlesec}
\titleformat{\section}[block]{\large\scshape\centering}{§\,\thesection}{1em}{}[\HRule{3pt}] 
\titleformat{\subsection}[block]{\normalsize\bfseries}{\thesubsection}{1em}{}
\titleformat{\subsubsection}[block]{\normalsize\bfseries}{\thesubsubsection}{1em}{}

\usepackage[framemethod=TikZ]{mdframed}

\newcommand{\beq}{\begin{equation}}
	\newcommand{\eeq}{\end{equation}}
\newcommand{\beqa}{\begin{eqnarray}}
	\newcommand{\eeqa}{\end{eqnarray}}
\newcommand{\CR}{\nonumber \\}

\newcommand{\m}{\mu}
\newcommand{\lam}{\lambda}

\usepackage[T1]{fontenc}

\newcommand{\HRule}[1]{\rule{\linewidth}{#1}} 	

\newcommand{\gq}{\mathfrak{q}}

\newcommand{\tens}{\otimes}

\newcommand{\inlim}{\mathop{\lim_{\longrightarrow}}}

\newcommand{\cals}[1]{\mathcal{#1}}

\newcommand{\bbs}[1]{\mathbb{#1}}

\newcommand*{\defeq}{\stackrel{\text{def}}{=}}
\newcommand{\normord}[1]{:\mathrel{#1}:}
\newcommand{\bb}[1]{\mathbb{#1}}

\numberwithin{equation}{section}
\newtheorem{thm}{Theorem}[section]

\newtheorem{lem}[thm]{Lemma}
\newtheorem{rem}[thm]{Remark}

\begin{document}
	
	\baselineskip = 18pt 
	
	\begin{titlepage}
		
		\bigskip
		\hfill\vbox{\baselineskip12pt
			\hbox{}
		}
		\bigskip
		\begin{center}
			\Large{ \scshape
				\HRule{3pt}
				Quasi-Hopf twist and elliptic Nekrasov factor 
				\HRule{3pt}
			}
		\end{center}
		\bigskip
		\bigskip
		
		\begin{center}
			\large 
			Panupong Cheewaphutthisakun$^{a}$\footnote{panupong.cheewaphutthisakun@gmail.com}
			and
			Hiroaki Kanno$^{a,b,}$\footnote{kanno@math.nagoya-u.ac.jp} \\
			\bigskip
			\bigskip
			$^a${\small {\it Graduate School of Mathematics, Nagoya University,
					Nagoya, 464-8602, Japan}}\\
			$^b${\small {\it KMI, Nagoya University,
					Nagoya, 464-8602, Japan}} \\
		\end{center}
		{\vskip 6cm}
		{\small
			\begin{quote}
				\noindent {\textbf{\textit{Abstract}}.}
				We investigate the quasi-Hopf twist of the quantum toroidal algebra of $\mathfrak{gl}_1$
				as an elliptic deformation. Under the quasi-Hopf twist the underlying algebra remains
				the same, but the coproduct is deformed, where the twist parameter $p$ is identified
				as the elliptic modulus. Computing the quasi-Hopf twist of the $R$ matrix, we 
				uncover the relation to the elliptic lift of the Nekrasov factor for instanton counting
				of the quiver gauge theories on $\mathbb{R}^4 \times T^2$. The same $R$ matrix 
				also appears in the commutation relation of the intertwiners, which implies
				an elliptic quantum KZ equation for the trace of intertwiners. We also show that 
				it allows a solution which is factorized into 
				the elliptic Nekrasov factors and the triple elliptic gamma function. 
		\end{quote}}

	\end{titlepage}
	
	
	\section{Introduction}
	
	The quantum toroidal algebra of type $\mathfrak{gl}_1$ (a.k.a. Ding-Iohara-Miki algebra \cite{DI}, \cite{Miki},
	the elliptic Hall algebra \cite{BS}, \cite{Sch}, \cite{SV} and others) is the fundamental symmetry which controls five dimensional 
	(or $K$ theoretic lift of) Nekrasov partition function \cite{Nekrasov:2002qd}, \cite{Nekrasov:2003rj},
	\cite{Nakajima:2003pg}, \cite{Nakajima:2005fg}. Its manifestation is
	the celebrated AGT relation \cite{AGT1}, \cite{AGT2}, \cite{AGT3}, \cite{Awata:2009ur}
	to the conformal block of deformed Virasoro and $W$ algebras. 
	The fact that the quantum toroidal algebra of type $\mathfrak{gl}_1$ has several names shows 
	its ubiquity and broad applications to many areas in mathematics and physics. 
	From the viewpoint of representation theory one of the advantages of the quantum toroidal
	algebra is that it has a coproduct which allows us to take the tensor product of the representations.
	In fact representations of the deformed Virasoro and $W$ algebras by free bosons 
	are derived from the tensor product of the Fock representations of the quantum toroidal algebra \cite{FHSSY}. 
	
	On the top of the hierarchy of supersymmetric gauge theories without coupling to the gravity is the six dimensional 
	theory. Recall that the BPS state counting of five dimensional theories on $\mathbb{R}^4 \times S^1$
	is naturally related to the supersymmetric quantum mechanics on the instanton moduli space, where $S^1$ 
	is identified with the (periodic) time direction. In the same manner
	the partition function of BPS state counting of the six dimensional theories on $\mathbb{R}^4 \times T^2$
	can be identified with the elliptic genus of the instanton moduli space \cite{Hollowood:2003cv}. 
	On the algebraic side we thus
	expect the appearance of the elliptic algebra and elliptic integrable systems \cite{Nekrasov:2013xda}. 
	In fact an elliptic lift of Ding-Iohara-Miki (DIM) algebra 
	together with its connection to the six dimensional Nekrasov function and the elliptic 
	Virasoro algebra has been already discussed in \cite{Nieri:2015dts}, \cite{Iqbal:2015fvd}.

	Most of the existing literatures employ 
	a version of elliptic DIM algebra that was first introduced by Y. Saito \cite{Saito} for the purpose 
	of describing an elliptic version of Macdonald polynomials. This kind of elliptic 
	algebra introduces a second set of deformed bosons, which is the same as the Clavelli-Shapiro
	method in old string theory \cite{Clavelli:1973uk} . The method allows us to rewrite the trace of the product
	of vertex operators (intertwiners) that appears in the one-loop diagrams in string theory
	by the vacuum expectation value with respect to the Fock vacuum of doubled boson system 
	\cite{Kimura:2016dys}, \cite{Ghoneim:2020sqi}.
	This also reminds us of the method of thermo field dynamics in statistical mechanics \cite{Takahasi:1974zn},
	\cite{Israel:1976ur}, \cite{Ojima:1981ma}. 
	However, there is another construction of an elliptic lift of DIM algebra, which is relatively
	unexplored \cite{FHHSY}\footnote{However, see \cite{Koroteev:2015dja} and \cite{Koroteev:2016znb}.}. 
	This construction does not employ an additional boson, 
	but makes use of the quasi-Hopf twist \cite{Jimbo:1999zz}, \cite{Jimbo:1998bi}. 
	In this paper we investigate the quasi-Hopf twist of DIM algebra.
	The original DIM algebra has two parameters $(q,t)$ and the quasi-Hopf twist introduces 
	a third parameter $p$. As we have emphasized the quantum toroidal algebra has a coproduct.
	The quasi-Hopf twist deforms the coproduct by what is called twistor $\cals{F}(p)$. The generating currents of 
	the algebra are also twisted accordingly. It turns out that the deformation parameter $p$ 
	is identified with an elliptic parameter. In the commutation relations of 
	the twisted currents there appears the theta function whose elliptic norm is $p$.

	As a quantum group the quantum toroidal algebra has an universal $R$ matrix \cite{FJMM}. 
	Since the quasi-Hopf twist deforms the coproduct, it also changes the universal $R$ matrix. 
	As has been shown in \cite{Awata:2016bdm}, \cite{Awata:2017cnz}, \cite{Awata:2017lqa},
	the Cartan part of the universal $R$ matrix is closely related 
	to the Nekrasov factor through the generalized Knizhnik-Zamolodchikov (KZ) equation for the correlation function of
	intertwiners of DIM algebra. Just like the Wick theorem for the free fields, the solutions to
	the generalized KZ equation are factorized into a product of two point functions.
	In accord with the AGT correspondence the Nekrasov factor plays 
	the role of the two point function \cite{Awata:2017cnz}, \cite{Cheewaphutthisakun:2021cud}.
	The elliptic Nekrasov factor is given by \cite{Nieri:2015dts}, \cite{Zhu:2017ysu}, \cite{Awata:2020yxf};
	\beqa\label{eNek}
	\mathcal{N}_{\lam \mu} (u \vert q, t, p)
	&=& \prod_{i, j =1}^\infty \frac{\Gamma(uq^{\lam_{j} -\mu_i} t^{i-j} ; q,p)}
	{\Gamma(u t^{i-j} ; q,p)}
	\cdot \frac{\Gamma (u t^{i-j+1} ; q,p)}
	{\Gamma (uq^{\lam_j - \m_{i}} t^{i-j+1} ; q,p)} \CR
	&=&
	\prod_{\square \in \lam} \theta_p (u q^{a_\lambda(\square)}  t^{\ell_\mu(\square)+1})
	\prod_{\square \in \mu} \theta_p (u q^{-a_\mu(\square) -1}  t^{-\ell_\lambda(\square)}),
	\eeqa
	where $(\lambda, \mu)$ is a pair of partitions and $(q,t) = (e^{\epsilon_1}, e^{-\epsilon_2})$ is
	the $\Omega$ background of Nekrasov \cite{Nekrasov:2002qd}. $\Gamma(u;q,p)$ is the elliptic gamma function. 
	In this paper we will show that the universal $R$ matrix after the quasi-Hopf twist is related 
	to the elliptic Nekrasov factor by the relation;
	\begin{equation}
		\gq^{| \lambda| + |\mu|} \mathcal{N}_{\lam \mu} (\gq^{-2} z \vert q, t, p)
		=\overline{R}_{\lambda\mu}(z ;p) \mathcal{N}_{\lam \mu} (z \vert q, t, p),
	\end{equation}
	where $\gq= \sqrt{t/q}$ and $\overline{R}_{\mu\lambda}$ is the normalized $R$ matrix of the vertical Fock representation.  
	With appropriate specialization of the spectral parameter $u$,
	the elliptic Nekrasov factor gives the contribution of the bifundamental matter hypermultiplet
	to the instanton partition function of the lift of $\mathcal{N}=2$ quiver gauge theory 
	to $\mathbb{R}^4 \times T^2$, where the elliptic modulus of the two dimensional torus $T^2$ 
	is identified with $p$. Hence, contrary to the case of \cite{Nieri:2015dts}, \cite{Iqbal:2015fvd},
	we do not have to introduce an additional boson (Heisenberg algebra) 
	to obtain the elliptic Nekrasov factor \eqref{eNek}. Only the quasi-Hopf twist suffices. 
	This is one of the main messages of the present paper. 
	The same normalized $R$ matrix also appears in the commutation relations of the intertwiners $\Psi_{\lambda}(v;p)$
	and the dual intertwiners $\Psi^*_{\lambda}(v;p)$;
	\begin{align}\label{intcom}
		\Psi_{\lambda}(v;p)\Psi_{\mu}(w;p)
		&= \frac{G_2( \frac{w}{v} ; p_*, q , t^{-1})}{G_2(\gq^{-2} \frac{v}{w} ; p_*, q , t^{-1})}
		\cdot \overline{R}_{\lambda\mu}\Big(\frac{v}{w};p_*\Big)
		\Psi_{\mu}(w;p)\Psi_{\lambda}(v;p),
		\\  \label{dintcom}
		\Psi^*_{\lambda}(v;p)\Psi^*_{\mu}(w;p)
		&= \frac{G_2(\frac{w}{v} ; p, q, t^{-1})}{G_2(\frac{v}{w} ; p, q , t^{-1})}
		\cdot \overline{R}_{\lambda\mu}\Big(\frac{v}{w};p\Big)^{-1}
		\Psi^*_{\mu}(w;p)\Psi^*_{\lambda}(v;p),
	\end{align}
	where $p_* = p \gq^{-2}$ and $G_2(u; p, q , t^{-1})$ is the double elliptic gamma function. 
	The ratio of the double elliptic gamma functions,
	which is independent of $\lambda$ and $\mu$, comes from the vacuum contribution. 
	Note that the elliptic parameter in $G_2$ and $\overline{R}_{\lambda\mu}$ 
	for the commutation relation of $\Psi_{\lambda}(v;p)$ is not $p$, but $p_*$.

	Based on the braiding relations \eqref{intcom}, \eqref{dintcom} and the cyclic property of the trace, 
	we can derive a difference equation for the trace of the product of intertwiners,
	\begin{equation}\label{trace-int}
		\mathrm{Tr} \left[ \widetilde{Q}^{d_1} Q^{d_2} 
		\Psi^{*}_{\mu_1}(w_1;p) \cdots \Psi^{*}_{\mu_n}(w_n;p) \Psi_{\lambda_1}(z_1;p) 
		\cdots \Psi_{\lambda_n}(z_n;p) \right],
	\end{equation}
	where $(d_1, d_2)$ is a pair of grading operaters of DIM algebra.
	The shift parameter of the difference equation is $Q$ and the $Q$-shift produces a product of the $R$ matrices $\overline{R}_{\lambda\mu}$
	(see section 6 for explicit forms). Thus, we can regard the $Q$-difference equation as a generalization of q-KZB equation for
	a genus one conformal blocks \cite{FR}, \cite{Felder:1994be}, \cite{FJMMN}. 
	As in the case of \cite{Awata:2017cnz}, \cite{Cheewaphutthisakun:2021cud}, 
	there is a solution whose building blocks are the Nekrasov factors. 
	A typical example of such building blocks looks like
	\begin{align}
		&G_3(z; p, Q, q, t^{-1}) \cdot G_3(Q z^{-1}; p, Q, q, t^{-1}) 
		\notag \\
		& \quad \prod_{k=0}^\infty \mathcal{N}_{\lambda \mu}( Q^k z \vert q, t^{-1}, p)
		\mathcal{N}_{\mu \lambda }( Q^{k+1} z^{-1}\vert q, t^{-1}, p),
	\end{align}
	where the triple elliptic gamma function $G_3$ represents the \lq\lq vacuum\rq\rq\ contribution. 
	In particular, when all the partitions are trivial in \eqref{trace-int}, the Nekrasov factors
	become also trivial and only $G_3$ factors, which are completely symmetric in four parameters  $(p, Q, q, t^{-1})$, survive.

	The fact that we keep the underlying algebra and only twist the coproduct seems to have the following 
	advantage. The DIM algebra has an $SL(2, \mathbb{Z})$ automorphism, called Miki automorphism \cite{Miki}. 
	This is an automorphism of the associative algebra, but not of the bialgebra. Hence, the Miki automorphism
	survives after the quasi-Hopf twist, though it is not clearly seen in terms of the twisted (elliptic) currents.
	On the other hand if we introduce an additional boson as in \cite{Saito}, the existence of 
	the Miki automorphism is not clear at all. Incidentally, we are also led to the following question;
	Since the Miki automorphism is not an automorphism of the bialgebra,
	it deforms the coproduct structure. Hence, one can ask if the change of the coproduct by the automorphism is 
	described by a twisting of the coproduct by an appropriate twistor.

	One of the interesting aspects of the quasi-Hopf twist of DIM algebra is the emergence of
	$SU(4)$ equivariant parameters. We have seen there appear two kinds of parameters $p$ and $p_*$
	in the commutation relations \eqref{intcom} and \eqref{dintcom}. More basically, as we will see in the next section, 
	in addition to the theta functions with the elliptic modulus $p$, the exchange relations of the twisted currents 
	involve those with the elliptic modulus $p_* = p C^{-2}$, where $C$ is one of the central charges
	of DIM algebra. 
	This is in sharp contrast with the elliptic DIM algebra defined in \cite{Saito}, where only the theta function
	with parameter $p$ appears. Recall that the original DIM algebra has parameters $(q_1, q_2, q_3)$
	with $q_1 q_2 q_3=1$. The standard Fock representation in terms of a free boson has the central
	charge $C=q_3^{1/2}$. After the quasi-Hopf twist with the twist parameter $p$,
	the Fock representation has $p_* = p q_3^{-1}= p q_1q_2$.
	Hence defining $q_3=p$ and $q_4 = p_*^{-1}$, we obtain $SU(4)$ parameters with $q_1 q_2 q_3 q_4=1$.
	We should not forget that these parameters are not associated with the algebra itself, but only arise in its Fock
	representation. However, the emergence of the $SU(4)$ parameters is quite suggestive.
	It is tempting to regard them as the equivariant parameters (or the $\Omega$ background) 
	of the torus action on $\mathbb{C}^4$, which plays the role of the ambient space of
	the spiked instanton (or the gauge origami) proposed by Nekrasov \cite{Nekrasov:2015wsu}. 
	It was introduced to provide a physical definition of the $qq$-character of 
	the $\mathcal{N}=2$ quiver gauge theories in terms of the brane configuration in type IIB string theory.

	\subsubsection*{Organization of material}
	The materials of this paper are structured as follows: in section \ref{section2QSHT} 
	we provide the definition of the quasi-Hopf twist of DIM algebra and the formula of its coproduct in terms of
	the twisted currents. 
	Since we keep the underlying associative algebra, representations of the original DIM algebra also work 
	as representations after the quasi-Hopf twist. We give corresponding representations in terms of the twisted currents
	in section \ref{section3reps}.  On the other hand, since the coproduct is deformed, the intertwiners will change.
	In section \ref{section4inter} after the intertwiner and the dual intertwiner are defined using the coproduct,
	we express them explicitly as operators on the Fock space of
	free bosons. We also provide formulas of the zero mode factor, which plays an important role in deriving 
	their commutation relations. The quasi-Hopf twist of the $ R $-matrix is derived from the universal $ R $-matrix
	of the DIM algebra in section \ref{section5Rmatrix}. 
	We double-check the computation by confirming that it agrees with the coefficient which results from
	interchanging the elliptic Fock intertwiners themselves, and dual elliptic Fock intertwiners themselves.  
	We also check the unitarity of the $R$ matrix and show a remarkable relation to the elliptic Nekrasov factor. 
	In section \ref{elliptic-qKZ} a difference equation for the trace of intertwiners and dual intertwiners is 
	derived from the cyclic property of the trace and the commutation relations among the intertwiners. 
	Some of technical details and auxiliary contents are relegated to Appendices.

	\subsection*{Elliptic functions}
	
	The complete odd theta function is defined by
	\begin{equation}
		\vartheta_1(z; p) := \sqrt{-1} \sum_{n \in \mathbb{Z}} (-1)^n p^{\frac{1}{2}\left(n- \frac{1}{2}\right)^2} z ^{n -\frac{1}{2}}.
	\end{equation}
	By the Jacobi triple product formula
	\begin{equation}
		\sum_{m \in \mathbb{Z}} p^{\frac{m^2}{2}} z^m 
		= \prod_{n=1}^\infty (1- p^n)(1+p^{n- \frac{1}{2}} z)(1+ p^{n - \frac{1}{2}}z^{-1}),
	\end{equation}
	we see
	\begin{equation}
		\sqrt{-1} (p;p)_\infty \cdot \theta_p(z) = p^{-\frac{1}{8}} z^{\frac{1}{2}} \vartheta_1(z; p),
	\end{equation}
	where we have defined a \lq\lq short\rq\rq\ theta function by
	\begin{equation}
		\theta_p(z) :=  (z;p)_\infty (pz^{-1};p)_\infty = \exp \left( - \sum_{n \neq 0} \frac{z^n}{n (1- p^n)} \right).
	\end{equation} 
	The \lq\lq short\rq\rq\ theta function enjoys the quasi-periodicity;
	\begin{equation}\label{theta-pshift}
		\theta_p(p^nz) = (-z)^{-n} p^{-\frac{1}{2}n(n-1)} \theta_p(z),
	\end{equation}
	and the inversion formula;	
	\begin{equation}\label{theta-inversion}
		\theta_p(z) = (-z)\theta_p(z^{-1}).
	\end{equation}
	We also use the elliptic gamma function;
	\begin{equation}
		\Gamma(z; q,p) := \frac{(qpz^{-1};q,p)_\infty}{ (z;q,p)_\infty} 
		= \exp \left( \sum_{n \neq 0} \frac{z^n}{n(1-q^n)(1- p^n)} \right).
	\end{equation} 
	The elliptic gamma function is symmetric in $q$ and $p$. It satisfies the $q$-difference equation;
	\begin{equation}\label{gamma-pshift}
		\Gamma(qz; q, p) = \theta_p(z) \Gamma(z; q, p).
	\end{equation}

	In general we can define an elliptic deformation of the multiple $q$-Pochhammer symbol by
	\begin{align}\label{eqPoch}
		(u;q_1,q_2, \cdots, q_n)_\infty &:= \exp \left( - \sum_{k=1}^\infty \frac{u^k}{k (1 - q_1^k)(1-q_2^k) \cdots (1-q_n^k)} \right) \CR
		&\longrightarrow  (u;q_1,\cdots, q_n, p)_\infty \cdot (q_1 \cdots q_n p u^{-1};q_1,\cdots, q_n, p)_\infty^{(-1)^n}.
	\end{align}
	In the literature \cite{Nishi} a multi-parameter generalization of the elliptic gamma function is defined by
	\begin{equation}
		G_n(u ; q_0, \cdots, q_n) := (u;q_0,\cdots, q_{r})_\infty^{(-1)^n} \cdot (q_0 \cdots q_{n} u^{-1};q_0, \cdots q_{r})_\infty,
	\end{equation}
	such that $G_0(u;q) = \theta_q(u), G_1(u;q_0, q_1) = \Gamma(u;q_0,q_1)$.
	Thus, the multiple elliptic gamma function $G_n(u; p, q_1, \cdots, q_n)^{(-1)^n}$ provides the elliptic lift of
	of the multiple  $q$-Pochhammer symbol \eqref{eqPoch}. 
	They satisfy the recursion relation
	\begin{equation}\label{gamma-rec}
		G_n(q_k u; q_0, \cdots, q_n) = 
		G_{n-1}(u;q_0, \cdots, \stackrel{k}{\vee}, \cdots, q_{n-1}) \cdot G_n (u;q_0, \cdots, q_{n}).
	\end{equation}
	The function $G_2(u ; q_0,q_1,q_2)$ is also called double elliptic gamma function.

	
	\section{Elliptic algebra from quasi-Hopf twist}
	\label{section2QSHT}
	
	\subsection{Quasi-Hopf twist of Ding-Iohara-Miki algebra}

	Let us begin with a review of the quantum toroidal algebra of type $ \mathfrak{gl}_1$, which we call
	Ding-Iohara-Miki (DIM) algebra in the present paper. The DIM algebra has the parameters $(q_1, q_2, q_3)$ with $q_1q_2q_3=1$
	and enjoys the triality of the permutation of $q_i$. We assume they are generic in the sense that for any $ a, b, c \in \mathbb{Z} $,
	\begin{gather}
		q^a_1q^b_2q^c_3 = 1 \implies a = b =c .
	\end{gather}
	We use the notation
	\begin{align}
		\kappa_n &:= \prod_{i=1}^3 (q_i^{\frac{n}{2}} - q_i^{-\frac{n}{2}}) = 
		\prod_{i=1}^3 (q_i^n -1) = \prod_{i=1}^3 (1- q_i^{-n}) = \sum_{i=1}^3 (q_i^n - q_i^{-n}),
	\end{align}
	which satisfies $\kappa_{-n} = - \kappa_n$. 
	By convention we often take $q=q_1$ and $t=q_2^{-1}$ as independent parameters. 
	It is convenient to introduce the notation $\gq := q_3^{\frac{1}{2}}= \sqrt{t/q}$ as the parameter of quantum deformation.

	We define the DIM algebra $\mathcal{U} := U_{q,t}(\widehat{\widehat{\mathfrak{gl}}}_1)$ 
	to be the associative algebra with the generators 
	$ E_k, F_k, K_0^{\pm}, H_r \,\, ( k \in \mathbb{Z}, r \in \mathbb{Z}\backslash\{	0	\})$ 
	and $ C $. It is convenient to introduce the generating functions (currents); 
	\begin{align}
		\label{DIMcurrents}
		E(z) = \sum_{k\in\mathbb{Z}} E_k z^{-k}, \quad
		F(z) = \sum_{k\in\mathbb{Z}} F_k z^{-k}, \quad
		K^{\pm}(z) = K_0^{\pm}\exp\left(
		\pm \sum_{r=1}^{\infty} H_{\pm r} z^{\mp r}
		\right).
	\end{align}
	There are several conventions of the Cartan currents $K^{\pm}(z)$. 
	The original convention is $K^{\pm}(z) = K^{\pm, \mathrm{here}}(C^{\frac{1}{2}}z)$.
	The advantage of our convention is that we can eliminate $C^{\frac{1}{2}}$ from the defining relations of the algebra. 
	Some literatures use the convention $K^{\pm}(z) = K^{\pm, \mathrm{here}}(Cz)$.
	
	The DIM algebra has two-dimensional center spanned by $(C,  K_0^{\pm})$. 
	Note that $K_0^{+}$ is the inverse of $ K_0^{-} $ by definition. 
	We will not write down the defining relations among the currents, since they can be recovered 
	from the relations after the quasi-Hopf twist by putting the deformation (elliptic) parameter $p=0$. 
	But we only quote the commutation relation 
	\begin{align}
		\label{DIMcartan} 
		[H_r, H_s]
		&= \delta_{r+s,0} \frac{\kappa_r}{r}(C^{r}-C^{-r}),
	\end{align}
	since it determines the normalization of the Cartan generators. 
	Actually there are also the Serre's relations in the defining relations, 
	which we do not write down explicitly, since they are not used in this paper.

	To define the quasi-Hopf twist of the DIM algebra, let us introduce the operators $ b_{\pm n} $ defined via
	\begin{align}\label{bmode} 
		K^+(z) = K^+_0\exp\bigg(\sum_{n=1}^{\infty}b_nC^{n}z^{-n}\bigg),
		\quad
		K^-(z) = K^-_0\exp\bigg(
		-\sum_{n=1}^{\infty}b_{-n}z^n
		\bigg).
	\end{align}
	The coproduct of $ b_{\pm n} $ is given by 
	\begin{align}
		\label{bcopro}
		\Delta(b_{\pm n}) = b_{\pm n} \otimes C_2^{-n} + 1 \otimes b_{\pm n}. 
	\end{align}
	Then in term of the twistor
	\begin{gather}\label{twistor}
		\mathcal{F}(p) = \exp\bigg(
		\sum_{n=1}^{\infty}\frac{np^nC^{-n}_2}{\kappa_n(1-p^nC^{-2n}_2)}
		b_n \tens b_{-n}
		\bigg) \in \mathcal{U} \tens \mathcal{U}, \qquad C_2 := 1 \tens C,
	\end{gather}
	we define the twisted coproduct by \cite{FHHSY}
	\begin{equation}\label{twistedcopro}
		\Delta_p(a) = \mathcal{F}(p)\Delta(a) \mathcal{F}(p)^{-1}, \qquad a \in \mathcal{U}.
	\end{equation}
	Note that the twistor $\mathcal{F}(p)$ is invertible\footnote{$\epsilon$ denotes the counit, which is 
		non-vanishing only on the central elements.};
	$(\epsilon \otimes \mathrm{id}) \mathcal{F}(p) =  (\mathrm{id} \otimes \epsilon ) \mathcal{F}(p) =1$. 
	By \eqref{bcopro} one can check that it satisfies the shifted cocycle condition \cite{FHHSY};
	\begin{align}
		\label{cocycle}
		\mathcal{F}^{(23)}(p)(\mathrm{id} \otimes \Delta) \mathcal{F}(p)
		= \mathcal{F}^{(12)}(pC_3^{-2})(\Delta\otimes \mathrm{id} ) \mathcal{F}(p)
	\end{align}
	on $\mathcal{U}^{\otimes 3}$. We would like to emphasize that by the quasi-Hopf twist 
	the underlying algebra $\mathcal{U}$ itself remains the same.
	But the coproduct has been deformed and hence the definition of tensor product representations will change. 
	Originally the DIM algebra $\mathcal{U}$ is a (quasi-triangular) Hopf algebra.
	But due to the deformation of the coproduct it is no longer true and it becomes a quasi-Hopf
	algebra (hence the name \lq\lq quasi-Hopf twist\rq\rq) with the Drinfeld associator;
	\begin{align}
		\Phi^{(D)} =  \mathcal{F}^{(23)}(p)(\mathrm{id} \otimes \Delta) \mathcal{F}(p) \cdot
		(\mathcal{F}^{(12)}(p)(\Delta\otimes \mathrm{id} ) \mathcal{F}(p))^{-1}.
	\end{align}
	For a quasi-Hopf algebra the coassociativity is modified by $\Phi^{(D)}$\footnote{For a quantum group
		the cocommutativity is broken, but the universal $R$ matrix compensates it.}. When $\Phi^{(D)}=1$
	the coassociativity holds and it is a Hopf algebra. Note that if there was no shift in the cocycle 
	condition \eqref{cocycle}, we had $\Phi^{(D)}=1$. Hence the shift of parameters in \eqref{cocycle}
	causes the violation of the coassociativity. On the other hand in the case of the elliptic DIM algebra 
	introduced in \cite{Saito}, the algebra is extended by an extra Heisenberg algebra, keeping the coproduct
	intact.

	\subsection{Coproduct and exchange relations among elliptic currents} 
	\label{sec2.3mainpaper}
	
	The new coproduct $\Delta_p$ is neatly expressed in terms of the elliptic currents to be defined shortly. 
	Introducing the twisting  currents by
	\begin{align}
		U^+(z;p) &= \exp\bigg(
		-\sum_{n=1}^{\infty}\frac{p^nC^{-n}}{1-p^nC^{-2n}}b_{-n}z^n
		\bigg),
		\\
		U^-(z;p) &=  \exp\bigg(
		\sum_{n=1}^{\infty}\frac{p^n}{1-p^n}b_nz^{-n}
		\bigg), 
	\end{align}
	we define the elliptic generating currents by 
	\begin{gather}
		E(z;p) = U^+(z;p)E(z), 	
		\qquad
		F(z;p) = F(z)U^-(z;p), 
		\label{mainpaper2.13}
		\\
		K^{+}(z;p) = U^+(z;p)K^+(z)U^-(C^{-1}z;p),
		\label{mainpaper2.14}
		\\
		K^-(z;p) = U^+(C^{-1}z;p)K^-(z)U^-(z;p).
		\label{mainpaper2.15} 
	\end{gather}
	Explicitly the twisted Cartan currents are
	\begin{align}
		K^{+}(z;p) &= K^{+}_0 \exp \left( - \sum_{n=1}^\infty \frac{p^n C^{-n}}{1- p^n C^{-2n}} b_{-n} z^n \right)
		\exp \left( \sum_{n=1}^\infty \frac{C^n}{1- p^n} b_{n} z^{-n} \right), \\
		K^{-}(z;p) &=  K^{-}_0 \exp \left( - \sum_{n=1}^\infty \frac{1}{1- p^n C^{-2n}} b_{-n} z^n \right)
		\exp \left( \sum_{n=1}^\infty \frac{p^n}{1- p^n} b_{n} z^{-n} \right).
	\end{align}
	Hence, they are related by the scaling of the spectral parameters;
	\begin{equation}\label{scaling}
		K^{+}(p^{-1} C z;p) = K^{+}_0 (K^{-}_0)^{-1} \cdot K^{-}(z;p). 
	\end{equation}
	Note that this relation ceases to hold for $p=0$, since it involves $p^{-1}$.

	In terms of the elliptic currents the coproduct is given by 
	\begin{align}
		\Delta_p\big(E(z;p)\big) &= E(z;p_*) \tens 1 + K^-(C_1z;p_*) \tens E(C_1z;p),
		\label{copro1}
		\\
		\Delta_p\big(F(z;p)\big) &= F(C_2z;p_*) \tens K^+(C_2z;p) + 1 \tens F(z;p),
		\\
		\Delta_p\big(K^+(z;p)\big) &= K^+(z;p_*) \tens K^+(C_1^{-1}z;p),
		\\
		\Delta_p\big(K^-(z;p)\big) &= K^-(C_2^{-1}z;p_*) \tens K^-(z;p),
		\label{copro4}
	\end{align}
	where $C_1 := C \tens 1, C_2 := 1 \tens C$ and $ p_* = pC_{2}^{-2} $.
	The deformed coproduct $\Delta_p$ becomes complicated in terms of the original currents.
	But the twisted currents makes it quite similar to the original copruduct $\Delta$.
	In fact $\Delta_p$ takes the same form as $\Delta$ except the shift of the elliptic parameter
	$p \to p_*$ in the first factor of the tensor product.
	Since the coproduct is a homomorphism; $\Delta_p(ab) = \Delta_p(a)\Delta_p(b)$, we can define the tensor product
	representation of $\rho_1$ and $\rho_2$ by $(\rho_1 \tens \rho_2) (a) = \rho_1 \tens \rho_2 (\Delta_p(a))$. 
	We would like to give a remark that the above coproduct is not a coproduct in the strict sense.
	The reason is that for each order $ k \in \bb{Z} $ of $ z^k $, 
	the expression of the coproduct contains an infinite summation of generators, which is not well-defined in general.

	Let us define an elliptic lift of the structure function of the DIM algebra by\footnote{The second equality is
		due to the condition $q_1 q_2 q_3 =1$.}
	\begin{gather}
		\cals{G}(x;p) := \frac{
			\theta_p(q_1^{-1}x)\theta_p(q_2^{-1}x)\theta_p(q_3^{-1}x)
		}{
			\theta_p(q_1x)\theta_p(q_2x)\theta_p(q_3x)
		}= \frac{
			\vartheta_1(q_1^{-1}x;p)\vartheta_1(q_2^{-1}x;p)\vartheta_1(q_3^{-1}x;p)
		}{
			\vartheta_1(q_1x;p)\vartheta_1(q_2x;p)\vartheta_1(q_3x;p)
		}.
	\end{gather}
	Using the relation \eqref{theta-pshift} and the property $q_1 q_2 q_3=1$,
	we can check that $\mathcal{G}(x; p)$ is periodic; $\mathcal{G}(px; p)=\mathcal{G}(x; p)$. 
	Similarly \eqref{theta-inversion} implies $\mathcal{G}(x^{-1}; p)=\mathcal{G}(x; p)^{-1}$. 
	Then the exchange relations of among elliptic currents can be stated as follows; 
	\begin{gather}
		K^{\pm}(z;p)K^{\pm}(w;p) = \frac{\cals{G}(w/z;p_*)}{\cals{G}(w/z;p)}K^{\pm}(w;p)K^{\pm}(z;p),
		\label{2.3main}
		\\
		K^{+}(z;p)K^-(w;p) = \frac{\cals{G}(w/\gq z;p_*)}{\cals{G}(\gq w/z;p)} K^-(w;p)K^{+}(z;p),
		\\
		K^+(z;p)E(w;p) = \cals{G}(w/z;p_*)E(w;p)K^+(z;p),
		\\
		K^-(\gq z;p)E(w;p) = \cals{G}(w/z;p_*)E(w;p)K^-(\gq z;p),
		\\
		K^+(\gq z;p)F(w;p) = \cals{G}(w/z;p)^{-1}F(w;p)K^+(\gq z;p),
		\\
		K^-(z;p)F(w;p) = \cals{G}(w/z;p)^{-1}F(w;p)K^-(z;p),
		\\ \label{EEexchage}
		E(z;p)E(w;p) = \cals{G}(w/z;p_*)E(w;p)E(z;p),
		\\
		F(z;p)F(w;p) = \cals{G}(w/z;p)^{-1}F(w;p)F(z;p),
		\\
		[E(z;p),F(w;p)] = \tilde{g}
		\bigg(
		\delta\Big(\frac{Cw}{z}\Big)K^+(z;p) - \delta\Big(\frac{Cz}{w}\Big)K^-(w;p) 
		\bigg),
		\label{2.11main}
	\end{gather}
	where the normalization factor $\tilde{g}$ of the commutation relation of $E(z;p)$ and  $F(z;p)$ does not
	change under the quasi-Hopf twist. Hence we can keep the same normalization $\tilde{g} = \kappa_1^{-1}$ 
	as \cite{Cheewaphutthisakun:2021cud}. On the other hand, the elliptic DIM algebra of \cite{Saito} choose a different normalization;
	the factor in $\tilde{g}$ is lifted to the theta functions. 
	Since we can change $\tilde{g}$ by the rescaling of $E(z;p)$ and  $F(z;p)$ without affecting other exchange relations, 
	the rescaling
	\beqa
	\label{scaleE}
	&&E(z,p) \longrightarrow \frac{(1-q_1)(q_1^{-1}p;p)_\infty (q_2^{-1};p)_\infty} {(p;p)_\infty (q_3 p;p)_\infty} E(z,p), \\
	\label{scaleF}
	&&F(z,p) \longrightarrow \frac{(1-q_1^{-1})(q_1 p;p)_\infty (q_2;p)_\infty} {(p;p)_\infty (q_3^{-1} p;p)_\infty} F(z,p).
	\eeqa
	is allowed for the matching of the normalization. Namely,
	by the rescaling \eqref{scaleE} and \eqref{scaleF}, we have
	\beq
	\frac{1}{\kappa_1} \longrightarrow 
	\frac{(q_1;p)_\infty (q_1^{-1}p;p)_\infty (q_2;p)_\infty (q_2^{-1} p;p)_\infty} {(p;p)_\infty^2 (q_3 p;p)_\infty (q_3^{-1};p)_\infty }
	= \frac{\theta_p(q_1) \theta_p(q_2)} {(p;p)_\infty^2 \theta_p(q_1q_2)},
	\eeq
	which exactly matches with the coefficient of the commutation relation (2) in \cite{WWYY}\footnote{The convention of the theta 
		function in \cite{WWYY} is different from ours.}.

	Since $\cals{G}(x;p)$ has infinitely many poles, it is mathematically precise to write the exchange relation \eqref{EEexchage}
	in the following way;
	\begin{align}
		& - \left( \frac{w}{z} \right)^3 
		\theta_{p_*}(q_1^{-1} \frac{z}{w})\theta_{p_*}(q_2^{-1} \frac{z}{w})\theta_{p_*}(q_3^{-1} \frac{z}{w}) \cdot E(z;p)E(w;p) 
		\notag \\
		&= \theta_{p_*}(q_1^{-1}\frac{w}{z})\theta_{p_*}(q_2^{-1}\frac{w}{z})\theta_{p_*}(q_3^{-1}\frac{w}{z}) \cdot E(w;p)E(z;p).
	\end{align}
	The same remark applies to other exchange relations.

	The elliptic parameter appearing in the exchange relations involving $F(z;p)$ is $p$,
	while it is the shifted parameter $p_*$ for $E(z;p)$\footnote{Since the power of $p_*$
		appears frequently, we have changed the original notation $p^*$ to $p_*$.}. 
	Note also that the relations \eqref{2.3main} -- \eqref{2.11main} are consistent with the scaling relation \eqref{scaling} of $K^{\pm}(z;p)$.
	In other words the relations involving $K^{-}(z;p)$ follow from those of $K^{+}(z;p)$. 
	When $C=1$ and hence $p_{*}=p$, these exchange relations agree with those of the elliptic DIM algebra
	introduced \cite{Saito} up to the normalization factor $\tilde{g}$ of the commutation relation $[ E(z;p), F(w;p)]$. 
	This in particular implies that the vertical representations with $C=1$ of Saito's elliptic algebra are also
	the vertical representations of the quasi-Hopf twist of the DIM algebra. 
	
	
	\section{Representations of the elliptic currents}
	\label{section3reps}
	
	Since the underlying algebra does not change as an associative algebra, the representations of the original DIM algebra 
	provide also those of the quasi-Hopf twisted algebra as representations of the associative algebra. 
	In particular there are central elements $(C, K_0^{-})$ which are constant, if the representation is irreducible. 
	Under the quasi-Hopf twist these values do not change. 
	Since only integer powers of $\gq$ appear as the values of the central elements in the present paper,
	we take the additive convention and define a representation has level $(n,m)$, if $(C, K_0^{-})= (\gq^n, \gq^{m})$. 
	On the other hand, the tensor product representations will change, since the coproduct is twisted. 
	As we will see this leads to an issue on the construction of the vertical Fock representation. 
	In this section we will express known representations of the original DIM algebra in terms of the elliptic currents.
	The advantage of using the twisted currents is that the coproduct $\Delta_p$ takes a simple form. 
	
	\subsection{Vector Representation}
	
	To obtain the vector representation of the elliptic currents, 
	we first start with the vector representation of the DIM algebra \cite{FFJMM}, and then perform the twisting procedure. 
	For each $ v \in \bb{C} $ called spectral parameter, let $ V(v) $ be the vector space over $ \bb{C} $ with a basis
	$ \{[v]_i|~i\in\bb{Z}\} $.  Recall that in the vector representation of the DIM algebra, we have
	\begin{equation}
		K^+(z)[v]_i = \widetilde{\psi}(q_1^{i}v/z)[v]_i,
		\qquad
		K^-(z)[v]_i = \widetilde{\psi}(q_1^{-i-1}z/v)[v]_i,
	\end{equation}
	where
	\begin{equation}
		\widetilde{\psi}(z) = \exp \left( \sum_{n=1}^{\infty}\frac{\kappa_n}{n}\frac{z^n}{1-q_1^{-n}} \right).
	\end{equation}
	Recall also that we define the operators $ b_{\pm n} $ by \eqref{bmode}.
	Since the vector representation has level $(0,0)$, we see
	\begin{equation}
		b_{\pm n} [v]_i = \frac{\kappa_n}{n}\frac{1}{1-q_1^{ \mp n}}(q_1^{i}v)^{\pm n} [v]_i.
	\end{equation}
	Hence, the twisting currents are given by 
	\begin{equation}
		U^+(z;p)[v]_i = \prod_{k=1}^{\infty}\widetilde{\psi}(p^kq_1^{-i-1}z/v)[v]_i,
		\qquad
		U^-(z;p)[v]_i = \prod_{k=1}^{\infty}\widetilde{\psi}(p^kq_1^iv/z)[v]_i. 
	\end{equation}	
	It is straightforward to check that in terms of the elliptic currents the vector representation $\rho_v^V$ is described by
	\begin{ceqn}
		\begin{align}
			K^+(z;p)[v]_i &= \frac{\theta_p(q_2^{-1}q_1^iv/z)\theta_p(q_3^{-1}q_1^iv/z)}{\theta_p(q_1^iv/z)\theta_p(q_1^{i+1}v/z)}[v]_i
			\\
			K^-(z;p)[v]_i &= \frac{\theta_p(q_3q_1^{-i}z/v)\theta_p(q_2q_1^{-i}z/v)}{\theta_p(q_1^{-i-1}z/v)\theta_p(q_1^{-i}z/v)}[v]_i, 
			\\
			E(z;p)[v]_i
			&= 
			\frac{
				(pq_2;p)_{\infty}(pq_3;p)_{\infty}
			}{
				(1-q_1)(p;p)_{\infty}(pq_1^{-1};p)_{\infty}
			}
			\delta\Big(
			q_1^{i+1}\frac{v}{z}
			\Big)
			[v]_{i+1},
			\\
			F(z;p)[v]_i
			&=
			\frac{
				(pq_2^{-1};p)_{\infty}(pq_3^{-1};p)_{\infty}
			}{
				(1-q_1^{-1})(p;p)_{\infty}(pq_1;p)_{\infty}
			}
			\delta\Big(
			q_1^{i}\frac{v}{z}
			\Big)
			[v]_{i-1}.
		\end{align}
	\end{ceqn}
	
	
	\subsection{Vertical Fock Representation by Tensor Product}
	\label{subsect3.2}
	
	As in the case of DIM algebra, we can construct the so-called vertical Fock representation 
	from the vector representation via the inductive limit \cite{FFJMM}. 
	The first step of this procedure is to perform the tensor product of vector representations
	with appropriate shift of spectral parameters;
	\begin{gather}
		V^n(v) \defeq V(v) \tens V(q_2v) \tens \cdots \tens V(q_2^{n-1}v). 
	\end{gather}
	For each $ \lambda = (\lambda_1, \dots, \lambda_n) \in \mathbb{Z}^n $, 
	we define $ |\lambda\rangle \in V^n(v) $ to be 
	\begin{gather}
		|\lambda\rangle \defeq 
		[v]_{\lambda_1-1} \otimes [q_2v]_{\lambda_2 -1 } \otimes \cdots \otimes
		[q^{n-1}_{2}v]_{\lambda_n - 1}. 
	\end{gather}
	It is clear that $ \{ |\lambda\rangle~| \lambda \in \bb{Z}^n\} $ forms a basis of $ V^n(v) $. We can endow the structure of
	$ \mathcal{U} $-module 
	$ \rho^{(n)} : \mathcal{U} \rightarrow \operatorname{End}(V^n(v))$
	to the vector space $ V^n(v) $ by\footnote{Since the coassociativity does not hold for $\Delta_p$, there is an ambiguity 
		in the definition of $\Delta_p^n$ in general. However, since $C=1$ for the vertical 
		representations, such a problem can be evaded. 
		We use the definition in \cite{Awata:2018svb} and \cite{Cheewaphutthisakun:2021cud}.}
	\begin{align}
		\rho^{(n)}(a)|\lambda\rangle =
		\Big[
		\rho^V_v \tens \rho^V_{q_2v} \tens \cdots \tens \rho^V_{q_2^{n-1}v}
		\Big]
		\Delta^{n-1}_p (a)|\lambda\rangle, \qquad a \in \cals{U}.
	\end{align}
	Note that here we can set $ C_1 = C_2 = 1 $ in the formulas 
	of the coproduct \eqref{copro1} -- \eqref{copro4}, 
	since we are focusing on the vector representations.
	For the elliptic currents the $ n $-fold tensor product of vector representations is given by
	\begin{align}
		\rho^{(n)}\big(K^+(z;p)\big)|\lambda\rangle &=
		\bigg[
		\prod_{i=1}^{n}\frac{\theta_p(q_1^{\lambda_i}q_2^{i}v/z)\theta_p(q_1^{\lambda_i - 1}q_2^{i-2}v/z)}
		{\theta_p(q_1^{\lambda_i}q_2^{i-1}v/z)\theta_p(q_1^{\lambda_i - 1}q_2^{i-1}v/z)}
		\bigg]
		|\lambda\rangle,
		\label{sss3.10}
		\\
		\rho^{(n)}\big(K^-(z;p)\big)|\lambda\rangle &=
		\bigg[
		\prod_{i=1}^{n}\frac{\theta_p(q_1^{-\lambda_i}q_2^{-i}z/v)\theta_p(q_1^{-\lambda_i + 1}q_2^{2-i}z/v)}
		{\theta_p(q_1^{-\lambda_i}q_2^{-i+1}z/v)\theta_p(q_1^{-\lambda_i + 1}q_2^{-i+1}z/v)}
		\bigg]
		|\lambda\rangle,
	\end{align}	
	and
	\begin{align}
		\rho^{(n)}\big(E(z;p)\big)|\lambda\rangle &= \sum_{k=1}^{n}
		\frac{
			(pq_2;p)_{\infty}(pq_3;p)_{\infty}
		}{
			(1-q_1)(p;p)_{\infty}(pq_1^{-1};p)_{\infty}
		}
		\delta\Big(
		q_1^{\lambda_k}q_2^{k-1}\frac{v}{z}
		\Big)
		\notag \\ 
		&~~\,\,\cdot
		\bigg[
		\prod_{i=1}^{k-1}\frac{\theta_p(q_1^{\lambda_k - \lambda_i}q_2^{k-i-1})\theta_p(q_1^{\lambda_k - \lambda_i +1}q_2^{k-i+1})}
		{\theta_p(q_1^{\lambda_k - \lambda_i}q_2^{k-i})\theta_p(q_1^{\lambda_k - \lambda_i + 1}q_2^{k-i})}
		\bigg]
		|\lambda + 1_k\rangle,
		\label{3.13mainpaper} 
		\\
		\rho^{(n)}\big(F(z;p)\big)|\lambda\rangle &=
		\sum_{k=1}^{n}
		\frac{
			(pq_2^{-1};p)_{\infty}(pq_3^{-1};p)_{\infty}
		}{
			(1-q_1^{-1})(p;p)_{\infty}(pq_1;p)_{\infty}
		}
		\delta\Big(q_1^{\lambda_k - 1}q_2^{k-1}\frac{v}{z}\Big)
		\notag \\
		&~~\,\,\cdot
		\bigg[
		\prod_{i = k+1}^{n}\frac{\theta_p(q_1^{\lambda_i-\lambda_k + 1}q_2^{i-k+1})\theta_p(q_1^{\lambda_i - \lambda_k}q_2^{i-k-1})}
		{\theta_p(q_1^{\lambda_i - \lambda_k + 1}q_2^{i-k})\theta_p(q_1^{\lambda_i - \lambda_k}q_2^{i-k})}
		\bigg]
		|\lambda - 1_k\rangle,
		\label{3.14mainpaper}
	\end{align}
	where $\lambda \pm 1_k$ means the shift of the $k$-th component $\lambda_k \to \lambda_k \pm 1$.
	Let us denote the set of partitions with length at most $n$ by 
	\begin{equation}
		\cals{P}_n := \left\{
		\begin{array}{@{}l@{}}
			\lambda = (\lambda_1,\cdots,\lambda_n) \in \bb{Z}^{n} \text{ s.t. }
			\lambda_1 \geq \cdots \geq \lambda_n \geq 0 
		\end{array} \right\}.
	\end{equation}
	For later convenience we also introduce the set of all partitions $\cals{P}$. 
	By the judicious choice of the $q_2$-shift of spectral parameters for a sequence of the vector representations,
	there is an invariant subspace 
	\begin{gather}
		W^{n,+}(v) \defeq 
		\text{span}\Big\{
		|\lambda\rangle \in V^n(v) \vert~\lambda \in \cals{P}_n 
		\Big\}.
	\end{gather}
	This can be confirmed by investigating the positions of zeros appearing in the action of the creation operator $E(z)$ 
	and the annihilation operator $F(z)$ \cite{FFJMM}.

	Next, we shall take the inductive limit of the tensor product of vector representations constructed above. 
	The reason why we are interested in taking the inductive limit is that we would like to remove the restriction 
	on the length of the partitions $\lambda$. 
	Thus, we consider the vector space $ \cals{F}_v $ which is defined by 
	\begin{align}
		\cals{F}_v\defeq \inlim W^{n,+}(v),
	\end{align}
	where the inductive limit is taken in the category of vector spaces.
	We would like to endow the structure of left
	$ \mathcal{U} $-module on $ \cals{F}_v $. 	
	At first glance, it is natural to define this representation 
	$ \rho^{\cals{F}} : \mathcal{U} \rightarrow \text{End}(\cals{F}_v) $ as follows ; 
	for each $ \lambda = (\lambda_1,\dots,\lambda_n,0,\dots) $ where $ \lambda_n \neq 0 $,
	\begin{gather}
		\rho^{\cals{F}} \big(X(z;p)\big)|\lambda\rangle \defeq 
		\rho^{(n+1)}\big(X(z;p)\big)|\lambda\rangle = \rho^{(n+2)}\big(X(z;p)\big)|\lambda\rangle = \cdots,
		\label{sss3.14}
	\end{gather}
	where $ X = K^{+}, K^-, E$ and $ F $. 
	That is, we expect that for all $ k \geq 1 $, $ \rho^{(n+k)}\big(X(z;p)\big)|\lambda\rangle $ are equal, 
	since the partitions $ (\lambda_1,\dots,\lambda_n), (\lambda_1,\dots,\lambda_n,0), (\lambda_1,\dots,\lambda_n,0,0), \dots  $ 
	are identified in the inductive limit. However, this is not the case, since
	from \eqref{sss3.10}, we see that for $ \lambda = (\lambda_1,\dots,\lambda_n,0,\dots) $ with $ \lambda_n \neq 0 $,
	\begin{gather}
		\frac{
			\langle \lambda|\rho^{(n+2)}\big(K^+(z;p)\big)|\lambda\rangle 
		}{
			\langle \lambda|\rho^{(n+1)}\big(K^+(z;p)\big)|\lambda\rangle 
		}
		= \frac{\theta_p(q_1^{-1}q_2^{n}v/z)\theta_p(q_2^{n+2}v/z)}{
			\theta_p(q_1^{-1}q_2^{n+1}v/z)\theta_p(q_2^{n+1}v/z) 
		}
		\neq 1.
	\end{gather}
	This means that we need certain modification factors in \eqref{sss3.14} for $ X = K^+ $.
	The same situation also occurs in the case $ X = K^- $. 
	Hence, let us define 
	\begin{gather}
		\rho^{\cals{F}} \big(K^+(z;p)\big)|\lambda\rangle 
		\defeq
		\beta_{n+1}\rho^{(n+1)}\big(K^+(z;p)\big)|\lambda\rangle
		= \beta_{n+2}\rho^{(n+2)}\big(K^+(z;p)\big)|\lambda\rangle
		= \cdots.
	\end{gather}
	Then we obtain a consistency condition for the modification factors $\beta_n$; 
	\begin{gather}
		\frac{\beta_{n+2}}{\beta_{n+1}}
		\frac{\theta_p(q_1^{-1}q_2^{n}v/z)\theta_p(q_2^{n+2}v/z)}{
			\theta_p(q_1^{-1}q_2^{n+1}v/z)\theta_p(q_2^{n+1}v/z) 
		}
		= 1. 
	\end{gather}
	Therefore, we conclude that $ \beta_n $ takes the following form; 
	\begin{gather}
		\beta_n = f(v/z)\frac{\theta_p(q_1^{-1}q_2^{n-1}v/z)}{\theta_p(q_2^{n}v/z)},
	\end{gather}
	where $ f(v/z) $ is a proportional factor which is independent of $ n $.

	By the same line of arguments, if we define for each $ \lambda = (\lambda_1,\dots,\lambda_n,0,\dots) $ with $ \lambda_n \neq 0 $,
	\begin{gather}
		\rho^{\cals{F}} \big(K^-(z;p)\big)|\lambda\rangle 
		\defeq
		\gamma_{n+1}\rho^{(n+1)}\big(K^-(z;p)\big)|\lambda\rangle
		= \gamma_{n+2}\rho^{(n+2)}\big(K^-(z;p)\big)|\lambda\rangle
		= \cdots. 
		\label{sss3.22}
	\end{gather}
	We then find that
	\begin{gather}
		\gamma_n = g(z/v)\cdot q_3\frac{\theta_p(z/q_1^{-1}q_2^{n-1}v)}{\theta_p(z/q_2^nv)}.
	\end{gather}
	From \cite{Awata:2018svb}, we expect that the modification factors $ \beta_n $ for $ K^+(z;p) $ 
	and $ \gamma_n $ for $ K^-(z;p) $ are the same. From this and the inversion formula
	\eqref{theta-inversion} we obtain that $f(v/z) = g(z/v)$.

	On the other hand, in the case $ X = E $, the problem does not arise. Namely,
	if $ \lambda = (\lambda_1,\dots,\lambda_n,0,\dots) $ with $ \lambda_n \neq 0 $, then the action
	\begin{gather}
		\rho^{\cals{F}} \big(E(z;p)\big)|\lambda\rangle \defeq 
		\rho^{(n+1)}\big(E(z;p)\big)|\lambda\rangle = \rho^{(n+2)}\big(E(z;p)\big)|\lambda\rangle = \cdots
	\end{gather}
	is well-defined. 
	From this result, we immediately see that we also have to introduce a modification factor 
	to the action of $ F(z;p) $, and, moreover, it has to be equal to $ \gamma_n $. That is,
	\begin{gather}
		\rho^{\cals{F}} \big(F(z;p)\big)|\lambda\rangle 
		\defeq
		\gamma_{n+1}\rho^{(n+1)}\big(F(z;p)\big)|\lambda\rangle
		= \gamma_{n+2}\rho^{(n+2)}\big(F(z;p)\big)|\lambda\rangle
		= \cdots. 
	\end{gather}
	The reason is that the defining relation \eqref{2.11main} must be satisfied.

	Finally, to accomplish the task, we have to determine an explicit expression of $ f(v/z) $.
	We require that the original Fock representation is recovered in the limit $ p \rightarrow 0 $.
	For simplicity, we also assume that $ f(v/z) $ does not depend on $ p $. 
	Thus, we conclude that 
	$f(v/z) = g(z/v) = \gq^{-1}$.

	As already mentioned, we can find the invariant subspace $\mathcal{F}_v$, which is spanned by the set of partitions $\cals{P}$.
	This is an irreducible subrepresentation generated by the empty partition $\varnothing$. We call it
	vertical Fock representation. It is a highest weight representation with the empty partition $\varnothing$ 
	being the highest weight state. In summary, we have constructed a representation 
	$ \rho^{\cals{F}} : \mathcal{U} \rightarrow \text{End}(\cals{F}_v) $ with the spectral parameter $ v $;
	\begin{ceqn}
		\begin{align}\label{verticalK+}
			\rho^{\cals{F}}  \big(K^+(z;p)\big)|\lambda\rangle &=
			\gq^{-1}
			\prod_{i=1}^{\infty}\frac{\theta_p(q_1^{\lambda_i}q_2^{i}v/z)\theta_p(q_1^{\lambda_i - 1}q_2^{i-2}v/z)}
			{\theta_p(q_1^{\lambda_i}q_2^{i-1}v/z)\theta_p(q_1^{\lambda_i - 1}q_2^{i-1}v/z)}
			|\lambda\rangle,
			\\
			\label{verticalK-}
			\rho^{\cals{F}} \big(K^-(z;p)\big)|\lambda\rangle &=
			\gq
			\prod_{i=1}^{\infty}\frac{\theta_p(q_1^{-\lambda_i}q_2^{-i}z/v)\theta_p(q_1^{-\lambda_i + 1}q_2^{-i+2}z/v)}
			{\theta_p(q_1^{-\lambda_i}q_2^{-i+1}z/v)\theta_p(q_1^{-\lambda_i + 1}q_2^{-i+1}z/v)}
			|\lambda\rangle,
			\\ 
			\rho^{\cals{F}} \big(E(z;p)\big)|\lambda\rangle &= 
			\sum_{k=1}^{\infty}
			\frac{
				(pq_2;p)_{\infty}(pq_3;p)_{\infty}
			}{
				(1-q_1)(p;p)_{\infty}(pq_1^{-1};p)_{\infty}
			}
			\delta\Big(
			q_1^{\lambda_k}q_2^{k-1}\frac{v}{z}
			\Big)
			\notag \\ 
			&~~\,\,\cdot
			\bigg[
			\prod_{i=1}^{k-1}\frac{\theta_p(q_1^{\lambda_k - \lambda_i}q_2^{k-i-1})\theta_p(q_1^{\lambda_k - \lambda_i +1}q_2^{k-i+1})}
			{\theta_p(q_1^{\lambda_k - \lambda_i}q_2^{k-i})\theta_p(q_1^{\lambda_k - \lambda_i + 1}q_2^{k-i})}
			\bigg]
			|\lambda + 1_k\rangle,
			\label{verticalE}
			\\ 
			\label{verticalF}
			\rho^{\cals{F}} \big(F(z;p)\big)|\lambda\rangle &= 
			\gq^{-1}
			\sum_{k=1}^{\infty}
			\frac{
				(pq_2^{-1};p)_{\infty}(pq_3^{-1};p)_{\infty}
			}{
				(1-q_1^{-1})(p;p)_{\infty}(pq_1;p)_{\infty}
			}
			\delta\Big(q_1^{\lambda_k - 1}q_2^{k-1}\frac{v}{z}\Big)
			\notag \\
			&~~\,\,\cdot
			\bigg[
			\prod_{i = k+1}^{\infty}\frac{\theta_p(q_1^{\lambda_i-\lambda_k + 1}q_2^{i-k+1})\theta_p(q_1^{\lambda_i - \lambda_k}q_2^{i-k-1})}
			{\theta_p(q_1^{\lambda_i - \lambda_k + 1}q_2^{i-k})\theta_p(q_1^{\lambda_i - \lambda_k}q_2^{i-k})}
			\bigg]
			|\lambda - 1_k\rangle. 
		\end{align}
	\end{ceqn}
	These are universal formulas which do not depend on the length of partition $\ell(\lambda)$. Our prescription 
	for the infinite product appearing in $K^{\pm}(z;p)$ and $F(z;p)$ is as follows; we make a successive 
	cancellation of the factors in the denominator and the numerator for $\lambda_n=0~ (\ell(\lambda) < n)$,
	which reduces the infinite product to a finite product once the partition $\lambda$ is fixed. 
	The factor $\gq^{\pm 1}$ is regarded as a result of the regularization of the infinite product by this prescription. 
	Note that it does not appear for $E(z;p)$ which does not require the infinite product. 
	Similarly the infinite sum for $E(z;p)$ and $F(z;p)$ reduces to a finite sum up to $\ell(\lambda)+1$, because when the adjacent lengths of the 
	partition agree; $\lambda_j= \lambda_{j+1}$, it is possible to have a factor $\theta_p(1)=0$. This also implies that
	when $ \lambda + 1_k $ or $ \lambda - 1_k $ is no longer a partition, the corresponding coefficient automatically vanishes. 
	From \eqref{verticalK-} we see that the vertical Fock representation has level $(0, 1)$.
	After the scaling \eqref{scaleE} and \eqref{scaleF} of $E(z;p)$ and $F(z;p)$, our result agrees with the vertical representation
	in \cite{WWYY}.

	\subsection{Vertical Fock Representation by Twisting}
	\label{subsect3.3}
	
	In the last subsection, we have constructed the vertical Fock representation by using the inductive limit 
	of the tensor product of vector representations. On the other hand, since the original DIM algebra has
	the vertical Fock representation \cite{FFJMM}, we may construct a vertical representation directly from the quasi-Hopf twist.
	In the vertical Fock representation, the operator $ b_n $ acts as follows:
	\begin{align}
		b_n|\lambda\rangle &= \frac{v^n}{n}\frac{\kappa_n}{1-q_1^{-n}}
		\bigg[
		\sum_{s=1}^{\ell(\lambda)}x^n_s + \frac{1}{1-q_2^n}x^n_{\ell(\lambda)+1}
		\bigg]|\lambda\rangle
		\notag \\
		&= \frac{v^n}{n}\frac{\kappa_n}{1-q_1^{-n}}\bigg[
		\sum_{s=1}^{\infty}x^n_s
		\bigg]
		|\lambda\rangle, \qquad  x_s := q_1^{\lambda_s -1} q_2^{s-1}. 
		\label{qqq2.20} 
	\end{align}
	We note that the eigenvalues of $b_n$ are proportional to $A_\lambda(q_1^n, q_2^n)$ to be defined below (see \eqref{equivch}). 
	Thus from \eqref{mainpaper2.13}, \eqref{mainpaper2.14} and \eqref{mainpaper2.15}, 
	we can check that the action of  the elliptic Cartan currents $ K^{\pm}(z;p)$ is the same as 
	\eqref{verticalK+} and \eqref{verticalK-}. On the other hand, we find some discrepancy 
	in the action of the elliptic currents $E(z;p)$ and $F(z;p) $. Namely there appear
	the following remainder factors against the formulas \eqref{verticalE} and \eqref{verticalF};
	\begin{align}
		R_{\lambda}^{(k)}(q_1, q_2 ;p)&:=
		\prod_{s=1}^{k-1}\frac{
			(pq_1^{\lambda_s - \lambda_k}q_2^{s-k};p)_{\infty}(pq_1^{\lambda_s - \lambda_k - 1}q_2^{s-k};p)_{\infty}
		}{
			(pq_1^{\lambda_s - \lambda_k}q_2^{s-k+1};p)_{\infty}(pq_1^{\lambda_s - \lambda_k - 1}q_2^{s-k-1};p)_{\infty}
		}
		\notag \\
		&\cdot 
		\prod_{s=k+1}^{\infty}\frac{
			(pq_1^{\lambda_k - \lambda_s + 1}q_2^{k-s+1};p)_{\infty}
			(pq_1^{\lambda_k - \lambda_s}q_2^{k-s-1};p)_{\infty}
		}{
			(pq_1^{\lambda_k - \lambda_s  + 1}q_2^{k-s};p)_{\infty}
			(pq_1^{\lambda_k - \lambda_s }q_2^{k-s};p)_{\infty}
		},
	\end{align}
	for $E(z;p)$ and 
	\begin{align}\label{tildeR}
		\widetilde{R}_{\lambda}^{(k)}(q_1, q_2 ;p)
		& = 	R_{\lambda - 1_k}^{(k)}(q_1, q_2 ;p)^{-1}
	\end{align}
	for $F(z;p)$. The relation \eqref{tildeR} allows us to understand 
	the remainders $R_{\lambda}^{(k)}$ and $\widetilde{R}_{\lambda}^{(k)}$ in the following manner; 
	In the vertical representations the Cartan currents $K^{\pm}(z;p)$ are commuting 
	and we employ a basis consisting of simultaneous eigenstates $\vert \lambda \rangle$ of $K^{\pm}(z;p)$. 
	Since the eigenvalues are non-degenerate, they are orthogonal. But there is an ambiguity 
	of the (relative) normalization of $\vert \lambda \rangle$, in particular it may depend on $\lambda$ and the elliptic parameter $p$. 
	The change of the normalization does not affect the matrix elements of $K^{\pm}(z;p)$, but
	the matrix elements of $E(z;p)$ and  $F(z;p)$ will change, since they are off-diagonal. In fact
	let us consider the change of the normalization;
	$\vert \lambda \rangle \longrightarrow \mathcal{C}_\lambda(q_1, q_2 ;p) \vert \lambda \rangle$,
	where $ \mathcal{C}_\lambda(q_1, q_2 ;p)$ is determined by the recursion relation
	\begin{equation}
		\frac{ \mathcal{C}_{\lambda +1_k} (q_1, q_2 ;p)}{ \mathcal{C}_\lambda(q_1, q_2 ;p)} = R_{\lambda}^{(k)}(q_1, q_2 ;p).
	\end{equation}
	Then one can see this change of the normalization eliminates both $R_{\lambda}^{(k)}$ and $\widetilde{R}_{\lambda}^{(k)}$.
	Finally with the initial condition $ \mathcal{C}_\varnothing (q_1, q_2 ;p)=1$,
	the recursion relation is solved by
	\begin{equation}
		\mathcal{C}_\lambda(q_1, q_2 ;p) = \prod_{s \in \lambda} ( pq_1^{a(s)} q_2^{-\ell(s)-1} ; p)_\infty.
	\end{equation}
	Here $ a(s) := \lambda_i -j $ and $ \ell(s) := \lambda^\vee_j -i $ are the arm-length and the leg-length 
	of the box $ s = (i,j) $ in the partition $ \lambda $. Note that
	\begin{equation}
		C_\lambda(q_1, q_2) := \prod_{s \in \lambda}( 1- q_1^{a(s)} q_2^{-\ell(s)-1})
	\end{equation}
	is the normalization factor which appears in the integral form of the Macdonald polynomials \cite{MacD}.

	\subsection{Horizontal Fock Representation}
	
	In subsection \ref{sec2.3mainpaper}, we have seen the relation between the elliptic currents 
	and the original DIM algebra, which has a Fock representation in terms of free bosons $\widetilde{a}_n$ \cite{FHHSY}. 
	The free boson operators obey the commutation relations of the deformed Heisenberg algebra;
	\begin{gather}
		[\widetilde{a}_n,\widetilde{a}_m] = \frac{n}{\kappa_n}(\gq^n - \gq^{-n})\delta_{n+m,0}
		= \frac{n \gq^{-n}}{(1-q_1^n)(1-q_2^n)}\delta_{n+m,0}.
	\end{gather}
	By using the relations \eqref{mainpaper2.13}--\eqref{mainpaper2.15}, 
	we can obtain the horizontal Fock representation of the elliptic currents.
	Let $ \cals{H}_u $ be a vector space over $ \bbs{C} $, which has
	\begin{gather}
		\left\{
		\widetilde{a}_{-\lambda_1}\cdots\widetilde{a}_{-\lambda_n}|0;u \rangle |
		\;\middle\vert\;
		\begin{array}{@{}l@{}}
			n\in\bbs{Z}^{\geq 0}, \\
			\lambda_1,\cdots,\lambda_n \in \bb{Z}^{>0} \text{ s.t. }
			\lambda_1 \geq \cdots \geq \lambda_n
		\end{array}
		\right\}
	\end{gather}
	as a basis. Here we have introduced the horizontal spectral parameter $u$ and the state $ |0; u\rangle $ is defined to be annihilated 
	by the positive mode operators $ \{\widetilde{a}_n|~n\in\bbs{Z}^{> 0}\} $. 
	
	If we define
	$ \rho^{(0)}_{H} : \mathcal{U} \rightarrow \text{End}(\cals{H}_u) $ by 
	\begin{align}
		\rho^{(0)}_{H}\Big(
		K^+(z;p)
		\Big)
		= &\exp\bigg(
		-\sum_{n=1}^{\infty}\frac{\kappa_n}{n}\frac{p^n\gq^{-3n/2}}{1-p^n_*}\widetilde{a}_{-n}z^n
		\bigg)
		\exp\bigg(
		\sum_{n=1}^{\infty}\frac{\kappa_n}{n}\frac{\gq^{n/2}}{1-p^n}\widetilde{a}_nz^{-n}
		\bigg),
		\\
		\rho^{(0)}_{H}\Big(K^-(z;p)\Big)
		= &\exp\bigg(
		-\sum_{n=1}^{\infty}\frac{\kappa_n}{n}\frac{\gq^{-n/2}}{1-p^n_*}\widetilde{a}_{-n}z^n
		\bigg)\exp\bigg(
		\sum_{n=1}^{\infty}\frac{\kappa_n}{n}\frac{p^n\gq^{-n/2}}{1-p^n}\widetilde{a}_{n}z^{-n}
		\bigg),
		\label{3.37mainreport}
		\\
		\rho^{(0)}_{H}\Big(	E(z;p)		\Big)
		=& \frac{u}{(1-q_1)(1-q_2)} \exp\bigg(
		\sum_{n=1}^{\infty}\frac{\kappa_n}{n}\frac{\gq^{-n/2}(1-p^n)}{(1-p^n_*)(\gq^n - \gq^{-n})}\widetilde{a}_{-n}z^n
		\bigg)
		\notag\\
		&\cdot\exp\bigg(
		-\sum_{n=1}^{\infty}\frac{\kappa_n}{n}\frac{\gq^{-n/2}}{\gq^n - \gq^{-n}}\widetilde{a}_{n}z^{-n}
		\bigg),
		\label{3.38mainreport}
		\\
		\rho^{(0)}_{H}\Big(	F(z;p) 	\Big)
		=
		& \frac{u^{-1}}{(1-q_1^{-1})(1-q_2^{-1})} \exp\bigg(
		-\sum_{n=1}^{\infty}\frac{\kappa_n}{n}\frac{\gq^{n/2}}{\gq^n - \gq^{-n}}\widetilde{a}_{-n}z^n
		\bigg) \notag\\
		&\cdot \exp\bigg(
		\sum_{n=1}^{\infty}\frac{\kappa_n}{n}\frac{\gq^{n/2}(1-p^n_*)}{(1-p^n)(\gq^n - \gq^{-n})}\widetilde{a}_nz^{-n}
		\bigg),
	\end{align}
	where $ p_* = p\gq^{-2} $. This is a level $(1,0)$ representation of $\mathcal{U} $.
	We can also obtain a level $(1, N)$ representation for any integer $ N $.
	By definition, the zero modes of $K^{\pm}(z;p)$ are $\gq^{\mp N}$. 
	The zero modes of $E(z;p)$ and $F(z;p)$ are fixed by consistency;
	\begin{equation}
		\mathbf{e}(z) \mathbf{f} (\gq^{-1}z) = \gq^{-N}, \qquad
		\mathbf{e}(\gq^{-1}z) \mathbf{f}(z) = \gq^N.
	\end{equation}
	Using a canonical solution
	\begin{equation}\label{zeromodes}
		\mathbf{e} (z)= \left( \frac{\gq}{z} \right)^N, \qquad
		\quad \quad
		\mathbf{f}(z) = \left( \frac{\gq}{z} \right)^{-N},
	\end{equation}
	we define the homomorphism $ \rho^{(N)}_{H}$;
	\begin{gather}
		\rho^{(N)}_{H}\Big(
		K^+(z;p)
		\Big)
		=
		\gq^{-N}
		\rho^{(0)}_{H}\Big(
		K^+(z;p)
		\Big),
		\quad \quad
		\rho^{(N)}_{H}\Big(
		K^-(z;p)
		\Big)
		=
		\gq^{N}
		\rho^{(0)}_{H}\Big(
		K^-(z;p)
		\Big),
		\notag \\ 
		\rho^{(N)}_{H}\Big(
		E(z;p)
		\Big)
		=
		\mathbf{e} (z)
		\rho^{(0)}_{H}\Big(
		E(z;p)
		\Big),
		\quad \quad
		\rho^{(N)}_{H}\Big(
		F(z;p)
		\Big)
		=
		\mathbf{f}(z)
		\rho^{(0)}_{H}\Big(
		F(z;p)
		\Big).
	\end{gather}
	In summary in the same way as the original DIM algebra, the horizontal Fock representations
	are characterized by the level $N$ and the spectral parameter $u$. 
	We will denote the free boson Fock space for the representation $  \rho^{(N)}_{H} $ by $\cals{H}_u^{(N)}$.

	\section{Intertwiner and Dual Intertwiner}
	\label{section4inter}
	
	In this section we construct the intertwining operator and the dual intertwining operator. 
	Historically they appeared 
	in the theory of solvable lattice models associated with the quantum affine algebra $U_q(\hat{\mathfrak{g}})$,
	where they were called vertex operators of type II and of type I, respectively \cite{MJ}.
	The vertical Fock representation $\cals{F}_v$ corresponds to the evaluation module in the case of the solvable lattice models
	and the horizontal Fock space $\cals{H}_u^{(N)}$ is a generalization of the level one 
	highest weight module of the quantum affine algebra. 
	The intertwining operator and the dual intertwining operator for the elliptic DIM algebra
	introduced by Y.Saito \cite{Saito} are constructed in \cite{Zhu:2017ysu}, \cite{Foda:2018sce}.	
	
	The intertwining operator $ \Psi (v;p) : \cals{F}_v \tens \cals{H}^{(N)}_u \rightarrow \cals{H}^{(N+1)}_w $ 
	is determined by the following intertwining condition \cite{Awata:2011ce};
	\begin{gather}\label{qqqintertwiner}
		a\Psi(v;p) = \Psi(v;p)\Delta_p(a), \quad a \in \mathcal{U}.
	\end{gather}
	Here $ \cals{F}_v $ denotes the vertical Fock representation that has level $(0,1)$, 
	while $ \cals{H}^{(N)}_u $ and $ \cals{H}^{(N+1)}_w $ are  horizontal representations of level $ (1,N) $ and $ (1,N+1) $, respectively. 
	Let $ \Psi (v) $ be the intertwiner defined by the original coproduct $\Delta$, which is given in \cite{Awata:2011ce}.
	Since the twisted coproduct $\Delta_p$ is defined by \eqref{twistedcopro},
	we see that schematically 
	\begin{equation}\label{twistint}
		\Psi (v;p) = \Psi (v)\cdot(\rho_{\cals{F}} \tens \rho_{H}) (\mathcal{F}(p)^{-1})
	\end{equation} 
	satisfies the condition \eqref{qqqintertwiner}. Recall that
	$ \{|\lambda \rangle\}_{\lambda \in \cals{P}}  $ forms a basis of $ \cals{F}_v $. 
	We define the $ \lambda $-component of the elliptic intertwiner $ \Psi_{\lambda} : \cals{H}^{(N)}_u \rightarrow \cals{H}^{(N+1)}_w$ by
	\begin{gather}
		\Psi_{\lambda}(v;p)\big(\bullet\big) = \Psi(v;p)\Big(|\lambda\rangle \tens \bullet\Big). 
	\end{gather}

	Similarly the dual intertwiner 
	$ \Psi^*(v;p) : \cals{H}^{(N)}_u \rightarrow \cals{H}^{(N-1)}_w \tens \cals{F}_v $
	is determined by the dual intertwining relation;
	\begin{gather}
		\Psi^*(v;p)a = \Delta_p (a)\Psi^*(v;p),\quad a \in \mathcal{U}. 
		\label{hainter}
	\end{gather}
	Again, from \eqref{twistedcopro}, if $\Psi^*(v)$ is the dual intertwiner before the quasi-Hopf twist, then 
	\begin{equation}
		\Psi^*(v;p) = (\rho_{H} \tens \rho_{\cals{F}}) (\mathcal{F}(p)) \cdot\Psi^*(v)
	\end{equation}
	gives a formal solution to the condition \eqref{hainter}.
	We define the $ \lambda $-component of the elliptic dual intertwiner 
	$ \Psi^*_{\lambda}(v;p) : \cals{H}^{(N)}_u \rightarrow \cals{H}^{(N-1)}_w   $ by 
	\begin{gather}
		\Psi^*(v;p)\Big(\bullet\Big) = \sum_{\lambda}\Psi^*_{\lambda}(v;p)\Big(\bullet\Big) \tens |\lambda\rangle . 
	\end{gather}

	\subsection{Elliptic Fock Intertwiner}
	
	For the intertwining relation \eqref{qqqintertwiner}, 
	we have $ C_1 = 1 $ and $ C_2 = \gq $ in \eqref{copro1} -- \eqref{copro4} 
	and the intertwining relations are explicitly;
	\begin{ceqn}
		\begin{align}
			K^+(z;p)\Psi_{\lambda}(v;p) &= \langle\lambda|K^+(z;p_*)|\lambda\rangle \Psi_{\lambda}(v;p)K^+(z;p),
			\label{4.1qqq}
			\\ 
			K^-(\gq z;p)\Psi_{\lambda}(v;p) &= \langle\lambda|K^-(z;p_*)|\lambda\rangle \Psi_{\lambda}(v;p)K^-(\gq z;p),
			\label{4.2qqq}
			\\
			E(z;p)\Psi_{\lambda}(v;p) &= 
			\sum_{k=1}^{\ell(\lambda) + 1}\langle\lambda+1_k|E(z;p_*)|\lambda\rangle \Psi_{\lambda+1_k}(v;p) +
			\langle \lambda|K^-(z;p_*)|\lambda\rangle \Psi_{\lambda}(v;p)E(z;p),
			\label{4.15mainpaper}
			\\
			F(z;p)\Psi_{\lambda}(v;p) &= \sum_{k=1}^{\ell(\lambda)}\langle\lambda - 1_k|F(\gq z;p_*)|\lambda\rangle\Psi_{\lambda - 1_k}(v;p)K^+(\gq z;p)
			+ \Psi_{\lambda}(v;p)F(z;p),
			\label{4.12qqq}
		\end{align}
	\end{ceqn}
	where $p_* = p \gq^{-2}$. 
	Note that here the inner product is calculated in the vertical representation $ \cals{F}_v $. 
	For the existence of the intertwiner the horizontal spectral parameters of the source 
	and the target Fock spaces have to be related by $w=-uv$ \cite{Awata:2011ce} (see also Appendix B). 
	We can express the elliptic Fock intertwiner $\Psi_{\lambda}(v;p)$ by the trivalent diagram in Figure \ref{figFockinter} below,
	where the change of the level and the spectral parameter of the horizontal Fock space is indicated. 
	\medskip
	\begin{figure}[h]
		\setlength{\unitlength}{1cm}
		\begin{center}
			\begin{picture}(10,3)
				\thicklines
				\put(10,1){\vector(-1,0){2.5}}
				\put(9.5,1.2){$ u $}
				\put(9,0.4){\small$(1,N)$}
				\put(2,1){\line(1,0){6}}
				\put(5,3){\vector(0,-1){1}}
				\put(5,1){\line(0,1){1}}
				\put(4.76,0.3){$  \lambda		$}
				\put(5.3,2){$ (0,1) $}
				\put(4.4,2){$  v $}
				\put(2,1){\vector(-1,0){0.5}}
				\put(0,1){\line(1,0){2}}
				\put(0,0.4){\small$(1,N+1)$}
				\put(0,1.2){$ -uv $}
			\end{picture}
		\end{center}
		\caption{Trivalent diagram of the elliptic Fock intertwiner $\Psi_{\lambda}(v;p)$}
		\label{figFockinter}
	\end{figure}
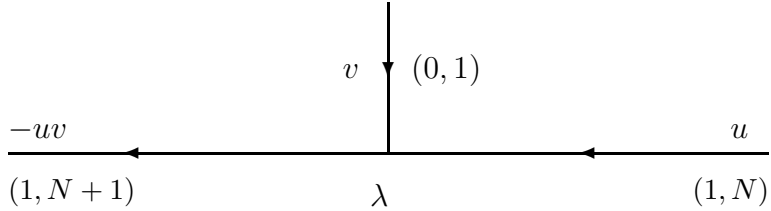
	
	\noindent The intertwining relations \eqref{4.1qqq} and \eqref{4.2qqq} mean $\Psi_{\lambda}(v;p)$ is an eigenstate 
	of the adjoint action of $K^+ (z;p)$ and $K^- (\gq z;p)$ with eigenvalues $\langle\lambda|K^\pm (z;p_*)|\lambda\rangle$. 
	These two conditions fix $\Psi_{\lambda}(v;p)$ up to the overall factor $z_{\lambda}(v)\cals{G}^{-1}_{\lambda}(p_*)$
	as follows;
	\begin{align}
		\Psi_{\lambda}(v;p) = z_{\lambda}(v)\cals{G}^{-1}_{\lambda}(p_*)
		&\exp\bigg(
		\sum_{n=1}^{\infty}\frac{1}{n}\frac{1-p^n}{1-p^n_*}\gq^{n/2}(q_2^n-1)(q_1^n - 1)
		\widetilde{a}_{-n}v^nA_{\lambda}(q_1^n,q_2^n)
		\bigg)
		\notag \\
		&\cdot 
		\exp\bigg(
		-\sum_{n=1}^{\infty}\frac{1}{n}\gq^{n/2}(q_2^n - 1)(q_1^n - 1)
		\widetilde{a}_nv^{-n} A_{\lambda}(q_1^{-n},q_2^{-n})
		\bigg),
		\label{4.19mainpaper}
	\end{align}
	where 
	\begin{equation}\label{equivch}
		A_{\lambda}(q_1,q_2) := \sum_{(i,j) \in \lambda}x_{ij} - \frac{1}{(q_1 - 1)(q_2 - 1)}, \qquad x_{ij} := q_1^{j-1} q_2^{i-1}
	\end{equation}
	is fixed by the eigenvalues $\langle\lambda|K^\pm (z;p_*)|\lambda\rangle$.
	The remaining two conditions \eqref{4.15mainpaper} and \eqref{4.12qqq} determine
	$z_{\lambda}(v)$ and $\cals{G}^{-1}_{\lambda}(p_*)$ as follows;
	\begin{equation}
		z_{\lambda}(v) = \prod_{(i,j) \in \lambda}
		(-\gq q_2^{i-1}x^{-1}_{ij}) u \cdot \mathbf{e}(x_{ij}v) = q_2^{n(\lambda)} (-v)^{-N \vert \lambda \vert} 
		u^{\vert \lambda \vert}f_\lambda(q_1, q_2)^{-N-1},
	\end{equation}
	and 
	\begin{equation}
		\cals{G}_{\lambda} (p_*) 
		= \prod_{s \in \lambda} \frac{(q_1^{-a(s)} q_2^{\ell(s)+1};p_*)_\infty}{(p q_1^{-a(s)} q_2^{\ell(s)+1};p_*)_\infty}.
	\end{equation}
	Here the framing factor $f_\lambda(q_1, q_2)$ is defined by 
	\begin{equation}\label{framing}
		f_{\lambda}(q_1,q_2) 
		\defeq \prod_{(i,j) \in \lambda} (-1) q_1^{j-1} q_2^{i-1} \gq^{-1}
		= \prod_{s \in \lambda}
		(-1)q_1^{a(s)+\frac{1}{2}}q_2^{\ell(s) + \frac{1}{2}}. 
	\end{equation}
	The intertwining relations of $E(z ;p)$ and $F(z ;p)$ are responsible for the formula of the zero mode factor $z_{\lambda}(v)$
	the normalization $\cals{G}_{\lambda}(p_*) $. More precisely, $z_{\lambda}(v)$ comes form the choice of the zero modes
	of the horizontal representation and $\cals{G}_{\lambda}(p_*) $ depends on the normalization of the basis of the vertical 
	Fock representation (See Appendices A and B for computations of $\cals{G}_{\lambda}(p_*) $ and $z_{\lambda}(v)$). 
	The appearance of $ \cals{G}_{\lambda}(p_*) $ and $ z_{\lambda}(v) $ can be described in the following manner.
	Since the vertical Fock representation is constructed as the semi-infinite product of 
	the vector representations (see subsec.\ref{subsect3.2}),
	we can express the elliptic Fock intertwiner \eqref{4.19mainpaper} as a composition of
	elliptic intertwiners for the vector representations \cite{Awata:2018svb}, 
	\cite{Cheewaphutthisakun:2021cud}. The factor $ \cals{G}_{\lambda}(p_*) $ is related to the normal ordering of this composition. 
	By the relation \eqref{twistint} the zero factor $ z_{\lambda}(v) $ is the same as the intertwiners for the original coproduct $\Delta$, 
	which are {\it e.g.} given in \cite{Awata:2018svb}.

	We note that $\Psi_{\lambda}(v;p)$ is expressed as a normal ordered product of the oscillator part $\eta(z;p)$ of the elliptic current $E(z;p)$;
	\begin{align}\label{etaproduct}
		\Psi_{\lambda}(v;p) &= z_{\lambda}(v) \cals{G}^{-1}_{\lambda}(p) : \Psi_{\varnothing}(v;p) 
		\prod_{(i,j) \in \lambda} \eta(q_1^{j-1} q_2^{i-1} v;p) :, \\
		\label{etaempty}
		& \Psi_{\varnothing}(v;p) := :\prod_{i.j=1}^\infty  \eta(q_1^{j-1} q_2^{i-1} v;p)^{-1}:.
	\end{align}
	The zero mode factor $z_\lambda(v)$, which depends on the level $N$ and the spectral parameters
	takes care of the zero mode factor of $E(z;p)$.
	This structure is exactly the the same as the intertwiner of the original DIM algebra, which is recovered by $p \to 0$,
	and explains the appearance of $A_{\lambda}(q_1,q_2)$
	which geometrically is the evaluation of the equivariant character of the tautological sheaf on the universal bundle of instantons
	at the fixed point labeled by $\lambda$. 
	
	It is remarkable that the shift of the spectral parameter $q_1^{j-1} q_2^{i-1} v$ in \eqref{etaproduct} and
	\eqref{etaempty} comes from the way of constructing the vertical Fock representation. In fact we have used 
	the matrix elements $\langle\lambda|K^\pm (z;p_*)|\lambda\rangle$ to fix this part.
	It is the vertical Fock representation that arises naturally from 
	the geometry of the Hilbert scheme of points on $\mathbb{C}^2$ \cite{Nak}. 
	In this way the Fock intertwiner incorporates the geometry of $U(1)$ instanton moduli space into the vertex operators
	on the Fock space of free bosons. To describe the moduli space of $U(N)$ instantons we have to take the $N$-fold
	tensor product of the boson Fock spaces.

	\subsection{Elliptic Dual Fock Intertwiner}
	
	In the case of the dual intertwining relation \eqref{hainter}, 
	$ C_1 = \gq, C_2 = 1 $ in \eqref{copro1} -- \eqref{copro4}. 
	Hence, the intertwining relations of the dual Fock intertwiner are explicitly;
	\begin{ceqn}
		\begin{align}
			\Psi^*_{\lambda}(v;p)K^+(z;p) &= \langle \lambda| K^+(\gq^{-1}z;p)|\lambda\rangle K^+(z;p)\Psi^*_{\lambda}(v;p),
			\label{6.3ja}
			\\
			\Psi^*_{\lambda}(v;p)K^-(z;p) &= \langle\lambda|K^-(z;p)|\lambda\rangle K^-(z;p)\Psi^*_{\lambda}(v;p),
			\label{6.4ja}
			\\
			\Psi^*_{\lambda}(v;p)E(z;p) &= E(z;p)\Psi^*_{\lambda}(v;p) + \sum_{k=1}^{\ell(\lambda)}
			\langle\lambda|E(\gq z;p)|\lambda - 1_k\rangle K^-(\gq z;p)\Psi^*_{\lambda-1_k}(v;p),
			\\
			\Psi^*_{\lambda}(v;p)F(z;p) &= \langle \lambda|K^+(z;p)|\lambda\rangle F(z;p)\Psi^*_{\lambda}(v;p) 
			+ \sum_{k=1}^{\ell(\lambda) + 1}\langle \lambda|F(z;p)|\lambda + 1_k\rangle\Psi^*_{\lambda + 1_k}(v;p).
			\label{4.21ja}
		\end{align}
	\end{ceqn}
	We can show that the solution of the intertwining relations \eqref{6.3ja} - \eqref{4.21ja} is
	\begin{align}
		\Psi^*_{\lambda}(v;p) = &z^*_{\lambda}(v) \cals{G}^{*-1}_{\lambda}(p)
		\exp\bigg(
		-\sum_{n=1}^{\infty}\frac{1}{n}\gq^{3n/2}(q_1^n - 1)(q_2^n - 1)\widetilde{a}_{-n}v^n
		A_{\lambda}(q_1^n,q_2^n)
		\bigg)
		\notag \\ 
		&\cdot 
		\exp\bigg(
		\sum_{n=1}^{\infty}\frac{1-p^n_*}{n(1-p^n)}(q_1^n-1)(q_2^n-1)\gq^{3n/2}\widetilde{a}_nv^{-n}
		A_{\lambda}(q_1^{-n},q_2^{-n})
		\bigg),
	\end{align}
	where 
	\begin{ceqn}
		\begin{align}
			z^*_{\lambda}(v) =& \gq^{|\lambda|}\prod_{(i,j) \in \lambda}
			(-\gq q_2^{i-1}x^{-1}_{ij})
			u^{-1}\cdot \mathbf{f}(x_{ij}v)
			= 
			\gq^{|\lambda|}q_2^{n(\lambda)}(-v)^{N|\lambda|}u^{-|\lambda|}
			f_{\lambda}(q_1,q_2)^{N - 1},
		\end{align}
	\end{ceqn}
	and 
	\begin{ceqn}
		\begin{align}
			\cals{G}^*_{\lambda}(p) =&
			\prod_{s \in \lambda}
			\frac{	\Big(	q_1^{a(s))}q_2^{-\ell(s)-1}			;p \Big)_{\infty}					}{
				\Big(p_*q_1^{a(s)}q_2^{-\ell(s)-1} ; p \Big)_{\infty}
			}. 
		\end{align}
	\end{ceqn}
	For the existence of the dual intertwiner the horizontal spectral parameters of the source 
	and the target Fock spaces have to be related by $w=-u/v$ \cite{Awata:2011ce}. 
	The elliptic dual Fock intertwiner $ \Psi^*_{\lambda}(v;p) $ is expressed 
	by the trivalent diagram in Figure \ref{figdualFockinter} below. 
	\begin{figure}[h]
		\setlength{\unitlength}{1cm}
		\begin{center}
			\begin{picture}(10,3)
				\thicklines
				\put(10,2.7){\vector(-1,0){2.5}}
				\put(9.5,2.9){$ u $}
				\put(9,2.1){\small$(1,N)$}
				\put(2,2.7){\line(1,0){6}}
				\put(5,2.7){\vector(0,-1){1}}
				\put(5,0.7){\line(0,1){1}}
				\put(4.76,0){$ \lambda $}
				\put(4.5,1.7){$ v $}
				\put(5.3,1.7){$ (0,1) $}
				\put(2,2.7){\vector(-1,0){0.5}}
				\put(0,2.7){\line(1,0){2}}
				\put(0,2.1){\small$(1,N - 1)$}
				\put(0,2.9){$ -u/v  $}
			\end{picture}
		\end{center}
		\caption{Trivalent diagram of the elliptic dual Fock intertwiner $ \Psi^*_{\lambda}(v;p) $}
		\label{figdualFockinter}
	\end{figure}
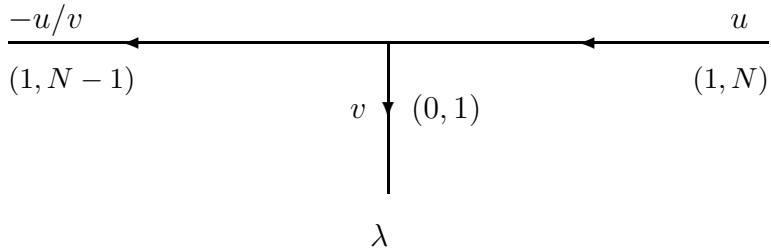
	
	We can observe a similarity to the case of the intertwiner. In fact we have
	\begin{align}
		\Psi^*_{\lambda}(v;p) &= z^*_{\lambda}(v) \cals{G}^{*-1}_{\lambda}(p) : \Psi^*_{\varnothing}(v;p) 
		\prod_{(i,j) \in \lambda} \xi(q_1^{j-1} q_2^{i-1} v;p) :, \\
		& \Psi^*_{\varnothing}(v;p) := :\prod_{i.j=1}^\infty  \xi(q_1^{j-1} q_2^{i-1} v;p)^{-1}:,
	\end{align}
	where $\xi(z;p)$ is the oscillator part of $F(z;p)$. 
	Namely, $E(z;p)$ for $\Psi_{\lambda}(v;p)$ is simply replaced by $F(z;p)$ for the dual intertwiner $\Psi^*_{\lambda}(v;p)$. 
	Again, the factor $ \cals{G}^{*-1}_{\lambda}(p) $ appears by removing the normal ordered product, when we express
	the dual intertwiner as a composition of those for the vector representation. Note also that
	the factor $ z^*_{\lambda}(v) $ depends on the level $ N $ of the horizontal Fock representation. 
	
	
	\section{Vertical $ R $-matrix and Elliptic Nekrasov Factor}
	\label{section5Rmatrix}
	
	In this section, we determine the $ R $-matrix corresponding to the vertical Fock representation.
	According to \cite{Jimbo:1999zz}, the quasi-Hopf twist of the universal $ R $-matrix $ \widetilde{\cals{R}} $ is
	\begin{gather}
		\widetilde{\cals{R}} = \cals{F}^{(21)}(p) \cdot \cals{R} \cdot \cals{F}^{-1}(p),
	\end{gather}
	where $ \cals{R} $ is the universal $ R $-matrix of the original DIM algebra and $\cals{F}(p)$ is the twistor given by 
	\eqref{twistor}. Here the notation $ \cals{F}^{(21)}(p) $ means that we interchange 
	the order of the elements in the tensor product of the expression of $ \cals{F}(p) $;
	\begin{gather}
		\cals{F}^{(21)}(p)
		= \exp\bigg(
		\sum_{n=1}^{\infty}\frac{np^nC^{-n}_1}{\kappa_n(1-p^nC^{-2n}_1)}
		b_{-n} \tens b_{n}
		\bigg). 
	\end{gather}

	According to \cite{FJMM}, the universal $ R $-matrix $ \cals{R}$ of DIM algebra factorizes as follows; 
	\begin{equation}
		\cals{R} = \gq^{-(c^\perp \otimes d^\perp + d^\perp \otimes c^\perp)} \cals{R}_{+} \cals{R}_{0} \cals{R}_{-},
	\end{equation} 
	where $\gq^{c^\perp}= K_0^{-}$ and $d^\perp= d_1$ (the grading operator for the principal degree).
	See also \cite{Garbali:2020sll} for computations of the $ R$ matrix for the horizontal Fock representation. 
	What is most relevant in the present paper is the Cartan factor $\cals{R}_{0}$
	with the contribution of the centers\footnote{The definition of $\kappa_n$ in this paper
		is $-\kappa_n$ in  \cite{FJMM}. We have changed the convention of the $R$ matrix from \cite{Cheewaphutthisakun:2021cud}.};
	\begin{gather}\label{universalR}
		\cals{R}_0^\prime = \gq^{-(c^\perp \tens d^\perp + d^{\perp}\tens c^{\perp}  ) }
		\exp\bigg(
		- \sum_{n=1}^{\infty}n\kappa_nh_{-n} \tens h_n
		\bigg),
	\end{gather}
	where $ h_{\pm n} $ is defined via $\kappa_n h_{\pm n} = \pm H_{\pm n}$.	
	Recall that any explicit formula of the universal $R$ matrix depends on the choice 
	of the Borel subalgebra from which the quantum group is reconstructed as the Drinfeld double. 
	It is interesting to find that the $ R $-matrix $ [\rho^{\cals{F}}_{v_1} \otimes \rho^{\cals{F}}_{v_2}] (\mathcal{R}_0) $, which appears shortly below, 
	coincides with the infinite slope $ R $-matrix $ R_\infty $ which is ubiquitous 
	in the Khoroshkin-Tolstoy factorization of the slope $ s $ $ R $-matrix introduced in \cite{Okounkov:2016sya}.
	From the viewpoint of the elliptic Hall algebra, $\cals{R}_0^\prime$ is a universal $ R$ matrix of the vertical (or slope infinity) Heisenberg subalgebra \cite{Negut:2020npc}.
	As noticed in \cite{Okounkov:2016sya}, $R_\infty$ corresponds to multiplication by a class of normal
	bundles in $K$-theory and is diagonal in the fixed point basis of the torus action.

	Since $c^{\perp}$ and $d^\perp$ commute with $b_{\pm}$, we obtain
	\begin{align}
		\widetilde{\cals{R}}_0^\prime &=  \gq^{-(c^\perp \tens d^\perp + d^{\perp} \tens c^{\perp} ) }
		\exp\bigg(
		\sum_{n=1}^{\infty}\frac{np^nC^{-n}_1}{\kappa_n(1-p^nC^{-2n}_1)}
		b_{-n} \tens b_{n}
		\bigg)
		\notag \\ 
		&\exp\bigg(
		-\sum_{n=1}^{\infty}n\kappa_nh_{-n} \tens h_n
		\bigg)
		\cdot\exp\bigg(
		-\sum_{n=1}^{\infty}\frac{np^nC^{-n}_2}{\kappa_n(1-p^nC^{-2n}_2)}
		b_n \tens b_{-n}
		\bigg).
		\label{5.5mainpaper}
	\end{align}
	The $ R $-matrix of the vertical Fock representation is evaluated as follows;
	\begin{align}
		\biggl\{
		[
		\rho^{\cals{F}}_{v_1} \tens \rho^{\cals{F}}_{v_2}
		]\Big(
		\widetilde{\cals{R}}_0^\prime
		\Big) 
		\biggr\}
		\Big(
		|\lambda,v_1\rangle \tens |\mu,v_2\rangle 
		\Big)
		=
		R_{\lambda\mu}\Big(\frac{v_1}{v_2};p\Big)
		\Big(
		|\lambda,v_1\rangle \tens |\mu,v_2\rangle 
		\Big),
	\end{align}
	where $ |\lambda,v_1\rangle \in \cals{F}_{v_1} $ and
	$ |\mu, v_2\rangle \in \cals{F}_{v_2} $. Since $c^{\perp}=1$ for the Fock representation and $d^{\perp}= d_1$ counts 
	the degree of the horizontal spectral parameter, we obtain
	\begin{align}
		R_{\lambda\mu}\Big(\frac{v_1}{v_2};p\Big) =
		& \gq^{- (| \lambda| + |\mu|)} \exp\bigg(
		\sum_{n=1}^{\infty}\frac{1}{1-p^n}\frac{v^{-n}_{1}}{1-q_1^n}
		\bigg[
		\sum_{s=1}^{\infty}x^{-n}_{s,\lambda_s}
		\bigg]
		\frac{v^n_2}{n}
		\frac{\kappa_n}{1-q_1^{-n}}
		\bigg[
		\sum_{r=1}^{\infty}x^n_{r,\mu_r}
		\bigg]
		\bigg)
		\notag \\ 
		&\cdot 
		\exp\bigg(
		-\sum_{n=1}^{\infty}\frac{p^n}{1-p^n}\frac{v^n_1}{1-q_1^{-n}}
		\bigg[
		\sum_{s=1}^{\infty}x^n_{s,\lambda_s}
		\bigg]
		\frac{v^{-n}_{2}}{n}\frac{\kappa_n}{1-q_1^n}
		\bigg[
		\sum_{r=1}^{\infty}x^{-n}_{r,\mu_r}
		\bigg]
		\bigg).
		\label{5.8mainpaper}
	\end{align}
	To simplify the expression \eqref{5.8mainpaper} the following formula can be used; 
	\begin{gather}
		\frac{1}{1- q_1}\bigg(\sum_{i=1}^{\infty} q_1^{\lambda_i} q_2^{i-1} \bigg) = - A_{\lambda}(q_1,q_2),
	\end{gather}
	where $A_{\lambda}(q_1,q_2)$ is defined by \eqref{equivch}. Then, we see that 
	\begin{align}
		R_{\lambda\mu}\Big(\frac{v_1}{v_2};p\Big) =
		& \gq^{- (| \lambda| + |\mu|)} \exp\bigg(
		\sum_{n=1}^{\infty}\frac{1}{1-p^n}\frac{\kappa_n}{n}\Big(\frac{v_2}{v_1}\Big)^n
		A_{\lambda}(q_1^{-n},q_2^{-n})A_{\mu}(q_1^n,q_2^n)
		\bigg)
		\notag \\ 
		&\cdot 
		\exp\bigg(
		-\sum_{n=1}^{\infty}\frac{p^n}{1-p^n}\frac{\kappa_n}{n}\Big(\frac{v_1}{v_2}\Big)^n
		A_{\mu}(q_1^{-n},q_2^{-n})A_{\lambda}(q_1^n,q_2^n)
		\bigg).
	\end{align}
	Let us define the normalized $ R $-matrix by
	\begin{equation}
		\overline{R}_{\lambda\mu} (z;p) := \frac{R_{\lambda\mu}(z;p) }{R_{\varnothing \varnothing}(z;p) },
	\end{equation}
	so that $\overline{R}_{\varnothing \varnothing}(z;p) =1$. 
	Then $\overline{R}_{\lambda\mu} (z;p)$ may also be expressed in terms of the theta function $ \theta_p(x) $
	by the following lemma:
	\begin{lem}
		\begin{align}
			\exp&\bigg(
			\sum_{n=1}^{\infty}\frac{1}{1-p^n}\frac{\kappa_n}{n}z^n
			A_{\lambda}(q_1^{-n},q_2^{-n})A_{\mu}(q_1^n,q_2^n)
			\bigg)
			\notag \\ 
			= &\prod_{(i,j) \in \lambda}\prod_{(k,l) \in \mu}
			\frac{
				\Big(zq_3^{-1}\frac{x_{kl}}{x_{ij}} ; p\Big)_{\infty}
				\Big(zq_2^{-1}\frac{x_{kl}}{x_{ij}} ; p\Big)_{\infty}
				\Big(zq_1^{-1}\frac{x_{kl}}{x_{ij}} ; p\Big)_{\infty}
			}{
				\Big(zq_3\frac{x_{kl}}{x_{ij}} ; p\Big)_{\infty}
				\Big(zq_2\frac{x_{kl}}{x_{ij}} ; p\Big)_{\infty}
				\Big(zq_1\frac{x_{kl}}{x_{ij}} ; p\Big)_{\infty}
			}
			\cdot \prod_{(i,j) \in \lambda}\frac{
				\Big(\frac{q_3z}{x_{ij}} ; p\Big)_{\infty}
			}{
				\Big(\frac{z}{x_{ij}} ; p\Big)_{\infty}
			}
			\notag \\ 
			&\cdot
			\prod_{(i,j) \in \mu}
			\frac{(zx_{ij};p)_{\infty}}{(zq_3^{-1}x_{ij};p)_{\infty}}
			\cdot 
			\exp\bigg(
			\sum_{n=1}^{\infty}\frac{1}{1-p^n}\frac{z^n}{n}\frac{1-q_3^{-n}}{(q_1^n - 1)(q_2^n - 1)}
			\bigg).
		\end{align}
	\end{lem}
	The result is
	\begin{align}
		\overline{R}_{\lambda\mu}\Big(\frac{v_1}{v_2};p\Big) =
		& \gq^{- (| \lambda| + |\mu|)} \prod_{(i,j)\in\lambda}\prod_{(k,l)\in\mu}
		\frac{
			\theta_p\Big(q_1^{-1}\frac{v_2}{v_1}\frac{x_{kl}}{x_{ij}}\Big)
			\theta_p\Big(q_2^{-1}\frac{v_2}{v_1}\frac{x_{kl}}{x_{ij}}\Big)
			\theta_p\Big(q_3^{-1}\frac{v_2}{v_1}\frac{x_{kl}}{x_{ij}}\Big)
		}{
			\theta_p\Big(q_1\frac{v_2}{v_1}\frac{x_{kl}}{x_{ij}}\Big)
			\theta_p\Big(q_2\frac{v_2}{v_1}\frac{x_{kl}}{x_{ij}}\Big)
			\theta_p\Big(q_3\frac{v_2}{v_1}\frac{x_{kl}}{x_{ij}}\Big)
		}
		\notag \\ 
		&\cdot
		\prod_{(i,j)\in\lambda}
		\frac{
			\theta_p(\frac{v_2}{v_1}\frac{q_3}{x_{ij}})
		}{
			\theta_p(\frac{v_2}{v_1}\frac{1}{x_{ij}})
		}
		\cdot 
		\prod_{(i,j)\in\mu}
		\frac{
			\theta_p(x_{ij}\frac{v_2}{v_1})
		}{
			\theta_p(q_1q_2x_{ij}\frac{v_2}{v_1})
		}.
	\end{align}
	By the inversion formula of the theta function \eqref{theta-inversion}, we can check 
	the unitarity of the normalized $ R $ matrix;
	\begin{equation}
		\overline{R}_{\mu\lambda}(z^{-1};p) = \overline{R}_{\lambda\mu}(z;p)^{-1}.
	\end{equation}
	
	We can derive the relation of the $ R $ matrix for the vertical Fock representation and the elliptic Nekrasov factor.
	Recall that we have obtained
	\begin{align}
		R_{\lambda\mu}(z;p) = 
		& \gq^{- (| \lambda| + |\mu|)} \exp\bigg(
		\sum_{n=1}^{\infty}\frac{1}{n}\frac{(1-q_2^{-n})(1-\gq^{-2n})}{(1-p^n)(1-q_1^n)}
		\sum_{i,j=1}^\infty \Big( x_{i, \lambda_i} x_{j, \mu_j}^{-1}z \Big)^{-n}
		\bigg)
		\notag \\ 
		&\cdot 
		\exp\bigg(
		-\sum_{n=1}^{\infty}\frac{1}{n}\frac{q_1^n p^n(1-q_2^n)(1-\gq^{2n})}{(1-p^n)(1-q_1^n)}
		\sum_{i,j=1}^\infty \Big( x_{i, \lambda_i} x_{j, \mu_j}^{-1}z \Big)^{n}
		\bigg) 
		\notag \\ 
		= & \gq^{- (| \lambda| + |\mu|)} 
		\prod_{i,j =1}^{\infty} \frac{\Gamma(q_1^{\mu_j - \lambda_i} q_2^{j-i}z^{-1}; q_1,p)
			\Gamma(\gq^{-2n}q_1^{\mu_j - \lambda_i} q_2^{j-i-1}z^{-1}; q_1,p)}
		{\Gamma(q_1^{\mu_j - \lambda_i} q_2^{j-i-1}z^{-1}; q_1,p)
			\Gamma(\gq^{-2n}q_1^{\mu_j - \lambda_i} q_2^{j-i}z^{-1}; q_1,p)}.
	\end{align}
	On the other hand the elliptic Nekrasov factor is (cf. \eqref{eNek});
	\begin{equation}
		\mathcal{N}_{\lam \mu} (u \vert q, t, p)
		= \prod_{i, j =1}^\infty \frac{\Gamma(uq^{\lam_{j} -\mu_i} t^{i-j} ; q,p)}
		{\Gamma (uq^{\lam_j - \m_{i}} t^{i-j+1} ; q,p)}\frac{\Gamma (u t^{i-j+1} ; q,p)}{\Gamma(u t^{i-j} ; q,p)}.
	\end{equation}
	Hence with $t=q_2^{-1}$ we see
	\begin{equation}\label{NandR}
		\frac{\mathcal{N}_{\lam \mu} (z \vert q, t, p)}{\mathcal{N}_{\lam \mu} (\gq^{-2} z \vert q, t, p)}
		= \gq^{| \lambda| + |\mu|} \frac{R_{\mu\lambda}(z^{-1} ;p)}{R_{\varnothing \varnothing}(z^{-1} ;p)} 
		= \gq^{| \lambda| + |\mu|} \overline{R}_{\mu\lambda}(z^{-1} ;p).
	\end{equation}
	
	By using the combinatorial identity (see Appendix E of \cite{Awata:2008ed})
	\begin{equation}
		\prod_{(i,j) \in \lambda} q^{\mu_i -j} \prod_{(i,j) \in \mu} q^{-\lambda_i + j-1} 
		= \prod_{(i,j) \in \mu} q^{\mu_i -j} \prod_{(i,j) \in \lambda} q^{-\lambda_i + j-1}
	\end{equation}
	and the inversion formula \eqref{theta-inversion},  we can prove
	\begin{equation}\label{Nek-inversion}
		\frac{\cals{N}_{\lambda\mu}(z\vert q_1,q_2,p)}{
			\cals{N}_{\mu\lambda}(\gq^{-2}z^{-1}\vert q_1,q_2,p)
		}
		= z^{|\mu| + |\lambda|}\gq^{|\mu|+|\lambda|}\frac{f_{\lambda}(q_1,q_2)}{f_{\mu}(q_1,q_2)},
	\end{equation}
	where the framing factor $f_\lambda(q_1, q_2)$ is defined by \eqref{framing}.
	The formula \eqref{Nek-inversion} also confirms the unitarity of the (normalized) $R$ matrix;
	\begin{ceqn}
		\begin{align}
			\overline{R}_{\mu\lambda}(z^{-1};p) 
			= \gq^{-|\mu|-|\lambda|} 
			\frac{\mathcal{N}_{\lambda\mu}(z\vert q_1,q_2,p)}{\mathcal{N}_{\lambda\mu}(\gq^{-2}z\vert q_1,q_2,p)}
			= \gq^{|\mu| + |\lambda|}
			\frac{
				\cals{N}_{\mu\lambda}(\gq^{-2}z^{-1}\vert q_1,q_2,p)
			}{
				\cals{N}_{\mu\lambda}(z^{-1}\vert q_1,q_2,p)
			}
			= \overline{R}_{\lambda\mu}(z;p)^{-1}.
		\end{align}
	\end{ceqn}

	We can also show that the Fock $ R $-matrix constructed above is obtained as
	the coefficient resulting from interchanging the intertwiners and the dual intertwiners themselves.
	The normal ordering of the vertex operators produces the elliptic Nekrasov factors. 
	Taking the difference of the zero mode factors into account and using \eqref{Nek-inversion}, we
	arrive at;
	\begin{ceqn}
		\begin{align}
			\Psi^*_{\mu}(w;p)\Psi_{\lambda}(v;p) &= \Upsilon\Big(\gq^{-1}\Big|\frac{v}{w};0\Big)
			\Psi_{\lambda}(v;p)\Psi^*_{\mu}(w;p),
			\label{5.13mainpaper}
			\\
			\Psi_{\lambda}(v;p)\Psi_{\mu}(w;p)
			&= \overline{R}_{\lambda\mu}\Big(\frac{v}{w};p_*\Big)
			\cdot\Upsilon(\gq^{-2}\vert \frac{v}{w};p_{*})
			\cdot \Psi_{\mu}(w;p)\Psi_{\lambda}(v;p),
			\label{5.14mainpaper}
			\\
			\Psi^*_{\lambda}(v;p)\Psi^*_{\mu}(w;p)
			&=
			\overline{R}_{\lambda\mu}\Big(\frac{v}{w};p\Big)^{-1}
			\Upsilon\left(1\Big\vert \frac{v}{w} ; p \right)
			\Psi^*_{\mu}(w;p)\Psi^*_{\lambda}(v;p),
			\label{5.15mainpaper}
		\end{align}
	\end{ceqn}
	where
	\begin{align}
		\Upsilon(\alpha | z; p) &\defeq
		\exp\bigg(
		\sum_{n=1}^\infty  \frac{1}{n}\frac{\alpha^n}{(1-p^n)(1-q_1^n)(1-q_2^n)}(z^n - z^{-n}) 
		\bigg)
		\notag \\ 
		&
		\qquad \exp\bigg(-
		\sum_{n=1}^\infty  \frac{1}{n}\frac{p_*^n \alpha^{-n}}{(1-p^n)(1-q_1^n)(1-q_2^n)}(z^n - z^{-n}) 
		\bigg)
		\notag \\ 
		&= \frac{G_2 (\alpha z^{-1} ; p, q_1, q_2)}{G_2 (\alpha z; p, q_1, q_2)}.
	\end{align}
	The equations \eqref{5.13mainpaper} -- \eqref{5.15mainpaper} are 
	elliptic generalization of eqs.(40) -- (42) in  \cite{Awata:2017cnz}.
	It is remarkable that the exchange relation between $\Psi_{\lambda}(v;p)$ and $\Psi^*_{\mu}(w;p)$ is 
	undeformed. The elliptic parameter for the exchange relation of the intertwiner is
	shifted $p \to p_*$.  Note also that
	\begin{equation}
		\Upsilon(\alpha | z^{-1} ; p) = \Upsilon(\alpha | z ; p)^{-1}.
	\end{equation}
	The emergence of the double elliptic gamma function $G_2$ is quite amusing, since it also appears 
	some of computations in six dimensional supersymmetric gauge theory (note that it is symmetric in $(p,q_1, q_2)$) 
	and topological strings \cite{Haghighat:2013gba}, \cite{Haghighat:2013tka}, 
	\cite{Lockhart:2012vp}, \cite{Lodin:2017lrc}, \cite{Mironov:2016cyq}.


	\section{Elliptic quantum Knizhnik-Zamolodchikov Equation}
	\label{elliptic-qKZ}
	
	In the elliptic case the quantum Knizhnik-Zamolodchikov (q-KZ) equation is a difference equation for the trace of intertwining operators
	\cite{EV}, \cite{ESV}, \cite{Konno:2017mos}, which is an analogue of the genus one conformal block of two dimensional conformal field theory;
	\begin{equation}
		\mathrm{Tr} \left[ q^{L_0} \phi_1(z_1) \cdots \phi_n(z_n) \right].
	\end{equation}
	Here $q= e^{2\pi i \tau}$ (not to be confused with the parameter of the DIM algebra), $\tau$ is the modulus of the 
	torus and $L_0$ is the zero mode of the Virasoro algebra. Contrary to the case of the vacuum expectation values 
	(matrix elements) of the product of intertwining operators, which corresponds to the genus zero conformal block
	on $\mathbb{P}^1$, the shift parameter is not fixed for the trace.

	Let us consider the trace of the intertwining operators\footnote{
	For simplicity in this section we will suppress the $p$-dependence of the intertwiners.
	};
	\begin{align}\label{tracedef}
		\mathfrak{G}_n(\vec{z}, \vec{w} \vert \vec{\lambda}, \vec{\mu}) 
		:= \mathrm{Tr}_{\cals{H}_u^{(N)}}  \left[ \widetilde{Q}^{d_1} Q^{d_2} 
		\Psi^{*}_{\mu_1}(w_1) \cdots \Psi^{*}_{\mu_n}(w_n) \Psi_{\lambda_1}(z_1) 
		\cdots \Psi_{\lambda_n}(z_n) \right],
	\end{align}
	where $d_1$ and $d_2$ are the grading operators that count the degree of the horizontal and the vertical parameters
	(see section 2.3 of \cite{Cheewaphutthisakun:2021cud}). In particular
	\begin{equation}
		\Psi_\lambda(Qz) =  Q^{d_2} \Psi_\lambda(z) Q^{-d_2}, \qquad \Psi^{*}_\mu(Qw) =  Q^{d_2} \Psi^{*}_\mu(w) Q^{-d_2}.
	\end{equation}
	Note that we can express the trace of the intertwining operators \eqref{tracedef} by the diagram in Figure \ref{tracetrivalentdiagram} below.
	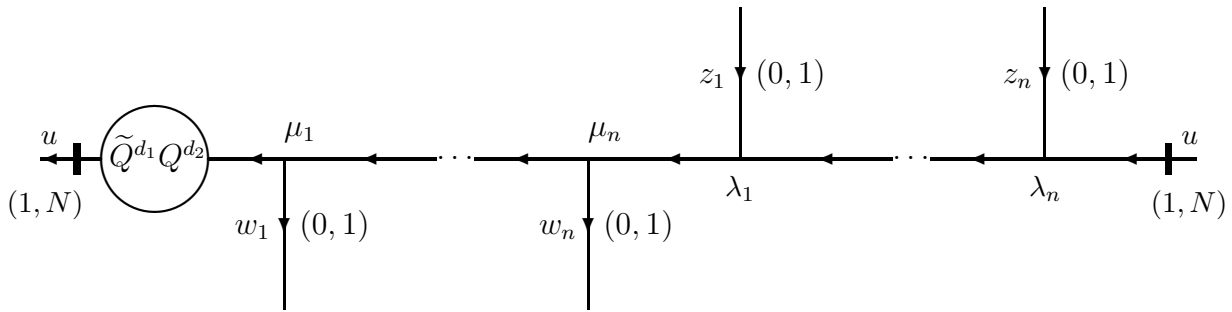
\begin{figure}[h]
		\setlength{\unitlength}{1cm}
		\begin{center}
			\begin{picture}(15,4)
				\thicklines
				\put(15,2){\vector(-1,0){1}}
				\put(13,2){\line(1,0){1}}
				\put(14.8,2.2){\small $ u $}
				\put(14.4,1.3){\small $ (1,N) $}
				
				\put(13,4){\vector(0,-1){1}}
				\put(13,2){\line(0,1){1}}
				\put(12.45,3){$ z_n $}
				\put(13.2,3){$ (0,1) $}
				\put(12.8,1.5){\small $ \lambda_n $}
				
				\put(13,2){\vector(-1,0){1}}
				\put(11.5,2){\line(1,0){0.5}}
				\put(11,2){$ \dots $}
				
				\put(11,2){\vector(-1,0){1}}
				\put(9,2){\line(1,0){1}}
				\put(9,4){\vector(0,-1){1}}
				\put(9,2){\line(0,1){1}}
				\put(8.45,3){$ z_1 $}
				\put(9.2,3){$ (0,1) $}
				\put(8.8,1.5){\small $\lambda_1$}
				
				\put(9,2){\vector(-1,0){1}}
				\put(7,2){\line(1,0){1}}
				\put(7,2){\vector(0,-1){1}}
				\put(7,0){\line(0,1){1}}
				\put(7,2.3){$ \mu_n $}
				\put(6.35,1){$ w_n $}
				\put(7.2,1){$ (0,1) $}
				
				\put(7,2){\vector(-1,0){1}}
				\put(5.5,2){\line(1,0){0.5}}
				\put(5,2){$ \dots $}
				
				\put(5,2){\vector(-1,0){1}}
				\put(3,2){\line(1,0){1}}
				\put(3,2){\vector(0,-1){1}}
				\put(3,0){\line(0,1){1}}
				\put(3,2.3){$ \mu_1 $}
				\put(2.35,1){$ w_1 $}
				\put(3.2,1){$ (0,1) $}
				
				\put(3,2){\vector(-1,0){0.5}}
				\put(2,2){\line(1,0){0.5}}
				\put(1.3,2){\circle{1.5}}
				\put(0.7,1.9){$ \widetilde{Q}^{d_1}Q^{d_2} $}
				\put(0.6,2){\vector(-1,0){0.8}}
				\put(-0.65,1.3){\small $ (1,N) $}
				\put(-0.2,2.2){\small $ u $}
				
				\put(0.3,1.8){\line(0,1){0.4}}
				\put(0.25,1.8){\line(0,1){0.4}}
				
				\put(14.65,1.8){\line(0,1){0.4}}
				\put(14.6,1.8){\line(0,1){0.4}}
				
			\end{picture}
		\end{center}
		\caption{Diagram representing the trace $\mathfrak{G}_n(\vec{z}, \vec{w} \vert \vec{\lambda}, \vec{\mu}) $.
			The left and the right ends of the horizontal line are identified. }
		\label{tracetrivalentdiagram}
	\end{figure}

	When we take the trace, the initial and the final Fock spaces must have the same level $N$ and the same spectral parameter $u$.
	Hence the number of the intetwiners and the dual intertwiners should agree. 
	Since we can shift the horizontal spectral parameter by $\widetilde{Q}^{d_1}$, we make the spectral parameters coincide
	by tuning the parameter $\widetilde{Q}$ as 
	\begin{equation}
		\widetilde{Q} = \prod_{i=1}^n \frac{w_i}{z_i}.
	\end{equation}
	Hence there remains a free parameter $Q$. We can derive a difference equation with the shift parameter $Q$
	as follows; let us first consider the case $z_k \to Q z_k$. By using \eqref{5.14mainpaper} we can move $\Psi_{\lambda_k} (Qz_k)$
	to the rightmost position in the trace, then by the cyclic property of the trace it is moved to the leftmost position. 
	Here it is important that this operation causes the change of the level and the spectral parameter of each intertwiner and dual intertwiner. 
	We need the compensating factor $\cals{C}$ associated with it. Then the commutation with $\widetilde{Q}^{d_1}$ adjusts 
	the zero mode factor of $\Psi_{\lambda_k} (Qz_k)$ appropriately by the scaling of the horizontal spectral parameter $u$ and
	the commutation with $Q^{d_2}$ cancels the $Q$-shift of $z_k$. Finally by using \eqref{5.13mainpaper} and \eqref{5.14mainpaper} again
	we can move $\Psi_{\lambda_k} (z_k)$ to the original position. Note that \eqref{5.13mainpaper} and \eqref{5.14mainpaper} already 
	take care of the change of the level and the spectral parameter associated with the exchange of the intertwiners. Consequently 
	no additional factors arise when we use them. After all these operations we arrive at 
	\begin{align}\label{Qdifference}
		Q^{z_k \frac{\partial}{\partial z_k}} \mathfrak{G}_n (\vec{z}, \vec{w} \vert \vec{\lambda}, \vec{\mu}) 
		=&~\cals{C} \cdot \prod_{j=1}^n\Upsilon \left(\gq^{-1} \vert \frac{z_k}{w_j};0 \right)^{-1} 
		\prod_{i<k} \overline{R}_{\lambda_i \lambda_k} \left( \frac{z_i}{z_k}; p_* \right)^{-1} 
		\Upsilon\left( \gq^{-2} \vert \frac{z_k}{z_i}; p_* \right)
		\notag \\
		& \quad \prod_{k<j} \overline{R}_{\lambda_k \lambda_j} \left( \frac{Q z_k}{z_j}; p_* \right)
		\Upsilon\left( \gq^{-2} \vert \frac{Q z_k}{z_j}; p_* \right) \cdot \mathfrak{G}_n (\vec{z}, \vec{w} \vert \vec{\lambda}, \vec{\mu}). 
	\end{align}
	Similarly in the case  $w_k \to Q w_k$, we can use \eqref{5.13mainpaper} and \eqref{5.15mainpaper} and obtain 
	\begin{align}\label{Qdifference2}
		Q^{w_k \frac{\partial}{\partial w_k}} \mathfrak{G}_n (\vec{z}, \vec{w} \vert \vec{\lambda}, \vec{\mu}) 
		=&~\cals{C}^{*} \cdot \prod_{i=1}^n \Upsilon \left(\gq^{-1} \vert \frac{z_i}{Q w_k};0 \right)
		\prod_{k<\ell} \overline{R}_{\mu_\ell \mu_k} \left( \frac{w_\ell}{Q w_k};p \right)
		\Upsilon\left( 1\vert \frac{Q w_k}{w_\ell}; p \right)
		\notag \\
		& \quad \prod_{\ell <k} \overline{R}_{\mu_k \mu_\ell} \left( \frac{w_k}{w_\ell}; p \right)^{-1}
		\Upsilon\left( 1\vert \frac{w_k}{w_\ell}; p \right) \cdot \mathfrak{G}_n (\vec{z}, \vec{w} \vert \vec{\lambda}, \vec{\mu}), 
	\end{align}
	with the factor $\cals{C}^{*}$ for the move of $\Psi^{*}_{\mu_k} (Qw_k)$ from the rightmost to the leftmost in the trace. 
	We can regard these $Q$-difference equations as a generalization of elliptic $q$-KZB equation to DIM algebra.

	\subsection{Computation of  $ \cals{C} $ }
	We define $ \cals{C} $ as follows : 
	\begin{ceqn}
		\begin{align}\label{calCdef}
			\mathrm{Tr} \bigg[
			&\widetilde{Q}^{d_1}Q^{-d_1}Q^{d_2}\Psi^*_{\mu_1}(w_1)\cdots\Psi^*_{\mu_n}(w_n)
			\Psi_{\lambda_1}(z_1)\cdots \stackrel{k}{\vee}
			\cdots\Psi_{\lambda_{n}}(z_n)\Psi_{\lambda_{k}}(Qz_k)
			\bigg]
			\notag \\ 
			&=  \cals{C} \cdot 
			\mathrm{Tr} \bigg[
			\widetilde{Q}^{d_1}Q^{d_2}
			\Psi_{\lambda_{k}}(z_k)
			\Psi^*_{\mu_1}(w_1)\cdots\Psi^*_{\mu_n}(w_n)
			\Psi_{\lambda_1}(z_1)\cdots \stackrel{k}{\vee}
			\cdots\Psi_{\lambda_{n}}(z_n)
			\bigg]. 
		\end{align}
	\end{ceqn}
	Recall that when $\Psi_\lambda(z)$ and $\Psi^{*}_\mu(w)$ act on the Fock space of level $N$ with the spectral parameter $u$, 
	their zero mode factors are
	\begin{equation}
		\Psi_\lambda(z) \sim (-z)^{-N|\lambda|} u^{|\lambda|} f_\lambda^{-N-1}, 
		\qquad
		\Psi^{*}_\mu(w) \sim (-w)^{N|\mu|} u^{-|\mu|} f_\mu^{N-1}.
	\end{equation}
	Chasing the change of the level and the spectral parameter, we find the total zero mode
	factor on the left hand side is
	\begin{ceqn}
		\begin{align}
			\cals{A}
			=& \bigg[	(-Qz_k)^{-N|\lambda_k|}u^{|\lambda_k|}f^{-N-1}_{\lambda_k}
			\bigg]
			\notag \\ 
			&\cdot
			\bigg[
			\prod_{i = k+1}^{n}
			(-z_i)^{ -(N + n - i +1)|\lambda_i|			}
			\Big(	u\cdot(-Qz_k)\prod_{j = i+1}^{n}(-z_j)				\Big)^{|\lambda_i|}
			f^{-N - 2 - n +i}_{\lambda_i}
			\bigg]
			\notag \\ 
			&\cdot 
			\bigg[
			\prod_{ i =1}^{k-1}
			(-z_i)^{-(N + n -i)|\lambda_i|}
			\Big( u\cdot(-Qz_k)\prod_{j = k+1}^{n}(-z_j)\prod_{l = i+1}^{k-1}(-z_l)						\Big)^{|\lambda_i|}
			f^{-N - n + i -1}_{\lambda_i}
			\bigg]
			\notag \\ 
			&\cdot 
			\bigg[
			\prod_{i = 1}^{n}
			(-w_i)^{	(N+i)|\mu_i|			}
			\Big(	\frac{	u\cdot(-Qz_k)\prod_{j \neq k}(-z_j)				}{	\prod_{j = i+1}^{n}(-w_j)		}					\Big)^{-|\mu_i|}
			f^{N + i - 1}_{\mu_i}
			\bigg],
		\end{align}
	\end{ceqn}
	where the initial level and spectral parameter of the horizontal Fock space are $N$ and $u$, respectively. 
	
	Next we are going to investigate the right hand side of \eqref{calCdef}.
	Similar consideration gives the total zero mode factor;
	\begin{align}
		\cals{B}
		=&
		\bigg[	(-z_k)^{-(N - 1)|\lambda_k|}				
		\Big(  \frac{	u \prod_{j \neq k}(-z_j)			}{\prod_{l = 1}^{n}(-w_l)}				\Big)^{|\lambda_k|}
		f^{-N}_{\lambda_k}
		\bigg]
		\notag \\ 
		&\cdot
		\bigg[
		\prod_{i = k+1}^{n}
		(-z_i)^{-(N + n - i)|\lambda_i|}
		\Big(	u \prod_{j = i+1}^{n}(-z_j)			\Big)^{|\lambda_i|}
		f^{-N - n + i - 1}_{\lambda_i}
		\bigg]
		\notag \\ 
		&\cdot 
		\bigg[
		\prod_{i = 1}^{k-1}
		(-z_i)^{-(N + n - i -1)|\lambda_i|}
		\Big(	u \prod_{i = k+1}^{n}(-z_i) \prod_{j = i+1}^{k-1}(-z_j)					\Big)^{|\lambda_i|}
		f^{-N -n+i}_{\lambda_i}
		\bigg]
		\notag \\ 
		&\cdot 
		\bigg[
		\prod_{i = 1}^{n}
		(-w_i)^{(N + i -1)|\mu_i|}
		\Big(	\frac{u	\prod_{j \neq k}(-z_j)		}{\prod_{l = i+1}^{n}(-w_l)}				\Big)^{-|\mu_i|}
		f^{N + i -2}_{\mu_i}
		\bigg].
	\end{align}
	After making the exchange with $Q^{d_2}$ which shifts all the vertical spectral parameters in $\cals{A}$ and $\cals{B}$;
	$Q^{d_2}\cals{A} = \cals{A}^{\prime}Q^{d_2}$and $Q^{d_2}\cals{B} = \cals{B}^{\prime}Q^{d_2}$,
	we obtain
	\begin{ceqn}
		\begin{align}
			\cals{C} = \frac{\cals{A}^{\prime}}{	\cals{B}^{\prime}	} =
			\frac{	\prod_{i = 1}^{n} w_i^{|\mu_i|}f_{\mu_i}			}{
				\prod_{i = 1}^{n}z^{|\lambda_i|}_{i}f_{\lambda_i}
			}
			\cdot 
			z_k^{ \sum_{i = 1}^{n}|\lambda_i| - \sum_{i=1}^{n} |\mu_i|					}\widetilde{Q}^{|\lambda_k|}
			\cdot 
			Q^{\sum_{i = 1}^{n}|\lambda_i| - \sum_{i=1}^{n} |\mu_i|				}
			Q^{-(N+1) |\lambda_k|}. 
		\end{align}
	\end{ceqn}

	\subsection{Computation of $ \cals{C}^* $}
	Now, we are going to compute the coefficient $ \cals{C}^* $ defined by
	\begin{ceqn}
		\begin{align}
			\label{2.1qdiff}
			\mathrm{Tr}
			&\bigg[
			\widetilde{Q}^{d_1}Q^{d_1}Q^{d_2}
			\Psi^*_{\mu_1}(w_1)\cdots \stackrel{k}{\vee}
			\cdots \Psi^*_{\mu_n}(w_n)
			\Psi_{\lambda_{1}}(z_1)\cdots \Psi_{\lambda_{n}}(z_n)\Psi^*_{\mu_k}(Qw_k)
			\bigg]
			\notag \\ 
			=& \cals{C}^* \cdot
			\mathrm{Tr}
			\bigg[
			\widetilde{Q}^{d_1}Q^{d_2}
			\Psi^*_{\mu_k}(w_k)\Psi^*_{\mu_1}(w_1)\cdots \stackrel{k}{\vee}
			\cdots \Psi^*_{\mu_n}(w_n)
			\Psi_{\lambda_{1}}(z_1)\cdots \Psi_{\lambda_{n}}(z_n)
			\bigg].
		\end{align}
	\end{ceqn}
	The total zero mode factor on the left hand side is
	\begin{align}
		\cals{M} =&
		\bigg[		(-Qw_k)^{N|\mu_k|}u^{-|\mu_k|}f^{N - 1}_{\mu_k}			\bigg]
		\notag \\ 
		&\cdot
		\bigg[
		\prod_{i = 1}^{n} (-z_i)^{	-(N - 1 + n -i)|\lambda_i|			}
		\Big(	-\frac{u}{Qw_k}\prod_{j = i+1}^{n}(-z_j)				\Big)^{|\lambda_i|}
		f^{-(N + n -i)}_{\lambda_i}
		\bigg]
		\notag \\ 
		&\cdot 
		\bigg[
		\prod_{ i = k+1}^{n}(-w_i)^{(N + i -1)|\mu_i|}
		\Big(	\frac{u	\prod_{j = 1}^{n}(-z_j)			}{(-Qw_k)\prod_{l = i+1}^{n}(-w_l)		}				\Big)^{-|\mu_i|}
		f^{N + i - 2}_{\mu_i}
		\bigg]
		\notag \\ 
		&\cdot 
		\bigg[
		\prod_{i = 1}^{k-1}(-w_i)^{(N + i)|\mu_i|}
		\Big(	\frac{u \prod_{j = 1}^{n}(-z_j)}{Q\prod_{l = i+1}^{n}(-w_l)}						\Big)^{-|\mu_i|}
		f^{N + i -1}_{\mu_i}
		\bigg].
	\end{align}

	On the other hand the zero mode factor on the right hand side of \eqref{2.1qdiff} is 
	\begin{align}
		\cals{L}
		=&
		\bigg[	(-w_k)^{	(N + 1)|\mu_k|		}
		\Big(	\frac{	u\prod_{j = 1}^{n}(-z_j)	}{	\prod_{l \neq k}(-w_l)				}				\Big)^{-|\mu_k|}
		f^{N}_{\mu_k}
		\bigg]
		\notag \\ 
		&\cdot 
		\bigg[
		\prod_{i = 1}^{n}(-z_i)^{	-(N + n -i)|\lambda_i|		}
		\Big(	u\prod_{j = i+1}^{n}(-z_j)				\Big)^{|\lambda_i|}
		f^{-(N + n - i) - 1}_{\lambda_i}D_{\lambda_i}
		\bigg]
		\notag \\ 
		&\cdot 
		\bigg[
		\prod_{i = k+1}^{n}
		(-w_i)^{(N + i)|\mu_i|}
		\Big(	\frac{u		\prod_{j = 1}^{n}(-z_j)	}{\prod_{l = i+1}^{n}	(-w_l)		}	\Big)^{-|\mu_i|}f^{N + i -1}_{\mu_i}
		\bigg]
		\notag \\
		&\cdot 
		\bigg[
		\prod_{i = 1}^{k-1} (-w_i)^{(N + i +1)|\mu_i|}
		\Big(	\frac{	u \prod_{j = 1}^{n}(-z_j)			}{	\prod_{j = i+1}^{k-1}(-w_j)\prod_{l = k+1}^{n}(-w_l)	}		\Big)^{-|\mu_i|}
		f^{N + i}_{\mu_i}
		\bigg].
	\end{align}	
	As before defining $\cals{M}^{\prime} = Q^{d_2}\cals{M}Q^{-d_2}$ and $\cals{L}^{\prime} = Q^{d_2}\cals{L}Q^{-d_2}$,
	we obtain
	\begin{align}
		\cals{C}^* = \frac{	\cals{M}^{\prime}		}{	\cals{L}^{\prime}			} =
		\frac{ \prod_{i = 1}^{n}z_i^{|\lambda_i|}f_{\lambda_i}					}{		
			\prod_{i = 1}^{n}w^{|\mu_i|}_{i}f_{\mu_i}
		}
		w_k^{	\sum_{i = 1}^{n}|\mu_i| - \sum_{i = 1}^{n}|\lambda_i|					}
		\widetilde{Q}^{-|\mu_k|}
		Q^{(N - 1)|\mu_k|}Q^{
			\sum_{i = 1}^{n}|\mu_i| - \sum_{i = 1}^{n}|\lambda_i| 
		}. 
	\end{align}
	
	\bigskip
	
	In \cite{Awata:2017cnz} a pair of the q-KZ equations for the trace of intertwiners 
	is derived and a solution is given rather explicitly (see eq.(84)). From the results we have derived in section \ref{section5Rmatrix}, 
	it is easy to guess a generalization of the solution in \cite{Awata:2017cnz}. 
	For example we can replace the Nekrasov factor (denoted as $G_{\alpha\beta}$ in \cite{Awata:2017cnz})
	by the elliptic one. To define a generalization of the remaining factor related to
	$\Upsilon(\alpha \vert z;p)$, let us introduce
	\beqa
	\widetilde{\Upsilon}(\alpha \vert z; p,Q) &:=&
	G_3(\alpha z ; p, Q, q_1, q_2) \cdot G_3(\alpha Q z^{-1} ; p, Q, q_1, q_2) \CR
	&=& \exp \left( \sum_{n=1}^\infty \frac{\alpha^n (z^n + Q^n z^{-n})}
	{n(1-p^n)(1-Q^n)(1-q_1^n)(1- q_2^n)} \right) \CR
	&&~~\exp \left( - \sum_{n=1}^\infty \frac{\alpha^{-n} p_*^n  (z^n + Q^n z^{-n})}
	{n(1-p^n)(1-Q^n)(1-q_1^n)(1- q_2^n)} \right).
	\eeqa
	Then the recursion relation \eqref{gamma-rec} among multiple elliptic gamma functions implies
	\beq\label{G3toG2}
	\widetilde{\Upsilon}(\alpha \vert Qz; p,Q) = \Upsilon(\alpha \vert z;p)^{-1}  \cdot \widetilde{\Upsilon}(\alpha \vert z; p,Q).
	\eeq

	We define the building blocks 
	\begin{align}
		\Theta_{\lambda \mu} (z \vert p, Q) :=& 
		\left( \prod_{k=0}^\infty \mathcal{N}_{\lambda \mu}(Q^k z \vert q_1, q_2, p) 
		\mathcal{N}_{\mu \lambda }(\gq^{-2} Q^{k+1} z^{-1} \vert q_1, q_2, p) \right)
		\widetilde{\Upsilon}( \gq^{-1} \vert \gq z; p,Q), 
		\notag \\
		\Phi_{\lambda \mu} (z \vert p, Q) :=& \left( \prod_{k=0}^\infty \mathcal{N}_{\lambda \mu}(\gq^{-2} Q^k z \vert q_1, q_2, p) 
		\mathcal{N}_{\mu \lambda }(\gq^{-2} Q^{k+1} z^{-1} \vert q_1, q_2, p) \right)
		\widetilde{\Upsilon}( \gq^{-2} \vert z; p,Q), 
		\\
		\overline{\Phi}_{\lambda \mu} (z \vert p, Q) :=& \left( \prod_{k=0}^\infty \mathcal{N}_{\lambda \mu}( Q^k z \vert q_1, q_2, p)
		\mathcal{N}_{\mu \lambda }( Q^{k+1} z^{-1}\vert q_1, q_2, p) \right)
		\widetilde{\Upsilon}( 1 \vert z; p,Q). \notag
	\end{align}
	It should be not an accident that, if we rewrite $\widetilde{\Upsilon}(\alpha \vert z; p,Q)$ in terms of the triple elliptic 
	gamma function $G_3(z ; p, Q, q_1, q_2)$, the argument $z$ agrees with those of $\mathcal{N}_{\lambda \mu}$ 
	and $\mathcal{N}_{\mu \lambda }$ with $k=0$.   Then our claim is that
	\beq\label{solution}
	\mathfrak{G}_n (\vec{z}, \vec{w} \vert \vec{\lambda}, \vec{\mu}) 
	= \cals{P} \cdot \frac{ \displaystyle{\prod_{i,j=1}^n} \Theta_{\lambda_i \mu_j} (\gq^{-1} z_i/w_j \vert 0, Q)}
	{ \displaystyle{\prod_{i< j}} \Phi_{\lambda_j \lambda_i } (z_j/z_i \vert p_*, Q) 
		\displaystyle{\prod_{k< \ell}} \overline{\Phi} _{\mu_\ell \mu_k} (w_\ell/w_k \vert p, Q)},
	\eeq
	where the monomial prefactor 
	\beq\label{prefactorQ}
	\cals{P} := \prod_{i=1}^n z_i^{- (N+n-i) |\lambda_i| - \sum_{i=1}^n |\mu_i| + \sum_{j<i} |\lambda_j|}
	\cdot w_i^{ (N+i) |\mu_i| + \sum_{j<i} |\mu_j|}
	\eeq
	is introduced for the matching of the $Q$ dependence of the factor $\cals{C}$ and $\mathcal{C}^{*}$ evaluated above. 
	The difference equation for \eqref{solution} is derived from those for the building blocks 
	$\Theta_{\lambda \mu}$, $ \Phi_{\lambda \mu}$ and $ \overline{\Phi}_{\lambda \mu} $.
	When the shift parameter is $Q$, by \eqref{G3toG2}, a direct computation shows 
	\beqa
	\frac{\Theta_{\lambda \mu} (Qz \vert p, Q)}{\Theta_{\lambda \mu} (z \vert p, Q) }
	&=& \mathcal{N}_{\lambda \mu} (z \vert q_1, q_2, p)^{-1}  
	\mathcal{N}_{ \mu \lambda} (\gq^{-2} z^{-1} \vert q_1, q_2, p)
	\Upsilon( \gq^{-1} \vert \gq z; p)^{-1} \CR
	&=& (\gq z) ^{-|\lambda| - |\mu|} (f_{\mu}/f_\lambda )\Upsilon( \gq^{-1} \vert \gq z; p)^{-1}, \\
	\frac{\Phi_{\lambda \mu} (Qz \vert p, Q)}{\Phi_{\lambda \mu} (z \vert p, Q) }
	&=& \mathcal{N}_{\lambda \mu} ( \gq^{-2} z \vert q_1, q_2, p)^{-1}  
	\mathcal{N}_{ \mu \lambda} ( \gq^{-2} z^{-1} \vert q_1, q_2, p)
	\Upsilon( \gq^{-2} \vert z; p)^{-1} \CR
	&=&  z^{-|\lambda| - |\mu|} (f_{\mu}/f_\lambda )  \overline{R}_{\lambda \mu} (z; p) ^{-1} 
	\Upsilon( \gq^{-2} \vert z; p)^{-1}, \\
	\frac{\overline{\Phi}_{\lambda \mu} (Qz \vert p, Q)}{\overline{\Phi}_{\lambda \mu} (z \vert p, Q) }
	&=& \mathcal{N}_{\lambda \mu} ( z \vert q_1, q_2, p)^{-1}  
	\mathcal{N}_{ \mu \lambda} ( z^{-1} \vert q_1, q_2, p)
	\Upsilon( 1 \vert z; p)^{-1} \CR
	&=&  z^{-|\lambda| - |\mu|} (f_{\mu}/f_\lambda ) \overline{R}_{\lambda \mu} (z; p) \Upsilon( 1 \vert z; p)^{-1}, 
	\eeqa
	where we have also used \eqref{NandR} and \eqref{Nek-inversion}. Hence, taking the power of $Q$ coming 
	from \eqref{prefactorQ} into account, we can see \eqref{solution} satisfies
	the difference equations \eqref{Qdifference} and \eqref{Qdifference2} with exactly the same factor $\cals{C}$ and $\cals{C}^{*}$.
	Let us explain the origin of these factors. 
	When we use the cyclic property of the trace 
	and move $\Psi_{\lambda_k} (z_k)$ or  $\Psi^{*}_{\mu_k}(w_k)$ from right to left, 
	it should be accompanied by the change of the level and the spectral parameter of the horizontal
	Fock space in the definition of the trace \eqref{tracedef}. 
	When we move $\Psi_{\lambda_k} (z_k)$, the change is
	$\mathrm{Tr}_{\cals{H}_u^{(N)}} \longrightarrow \mathrm{Tr}_{\cals{H}_{-z_ku}^{(N+1)}}$.
	On the other hand, when we move $\Psi^{*}_{\mu_k}(w_k)$ it is
	$\mathrm{Tr}_{\cals{H}_u^{(N)}} \longrightarrow \mathrm{Tr}_{\cals{H}_{-u/w_k}^{(N-1)}}$.
	We can check the total changes of the zero modes are exactly given by  $\cals{C}$ and $\cals{C}^{*}$
	up to the additional powers of $Q$. Finally the power of $\widetilde{Q}$ comes from the exchange with $\widetilde{Q}^{d_1}$.

	Since the coupled equations \eqref{Qdifference} and \eqref{Qdifference2} are $Q$-difference equations,
	there is an ambiguity of  \lq\lq integration constant\rq\rq\ or $Q$-periodic function in general. Let us illustrate this point in the simplest
	example of $n=1$ with empty partitions;
	\begin{equation}
		\mathfrak{G}_1 (z, w \vert \varnothing , \varnothing) 
		= \exp \left( \sum_{n=1}^\infty \frac{1}{n} \frac{(z/w)^n + Q^n \gq^{-2n} (w/z)^n}{(1-Q^n)(1-q_1^n)(1-q_2^n)}  \right).
	\end{equation}
	Unfortunately the solution is $p$ independent in this case. The parameter $Q$ is identified with
	the gauge coupling (the parameter of instanton expansion), which is consistent with AGT dictionary \cite{AGT1}.  
	Hence the substitution of $Q=0$ gives the perturbative part;
	\begin{equation}
		\mathfrak{G}_1^{\mathrm{pert}}
		= \exp \left( \sum_{n=1}^\infty \frac{1}{n} \frac{(z/w)^n}{(1-q_1^n)(1-q_2^n)}  \right).
	\end{equation}
	Then the instanton part is 
	\begin{equation}\label{n1inst}
		\mathfrak{G}_1^{\mathrm{inst}} =  \mathfrak{G}_1 / \mathfrak{G}_1^{\mathrm{pert}}
		= \exp \left( \sum_{n=1}^\infty \frac{1}{n} \frac{Q^n ((z/w)^n + \gq^{-2n}(w/z)^n)}{(1-Q^n)(1-q_1^n)(1-q_2^n)}  \right).
	\end{equation}
	On the other hand in this case we can compute the trace directly \cite{Iqbal:2008ra}, \cite{Poghossian:2008ge}, \cite{Awata:2009yc}, 
	\cite{Rains:2016xon}, \cite{Fukuda:2020czf}. For example, we quote the formula from \cite{Carlsson:2013jka};
	\begin{equation}\label{U1adj}
		Z^{\mathrm{inst}} (m, Q ; q_1, q_2) = \exp \left( \sum_{n=1}^n \frac{Q^n}{n m^n} 
		\frac{(m^n - q_1^n)(m^n - q_2^n)}{(1- Q^n)(1 - q_1^n)(1 - q_2^n)}  \right),
	\end{equation}
	where $m$ is the (exponentiated) mass of the adjoint matter hypermultiplet of $\cals{N} =2^{*}$ $U(1)$ gauge theory. 
	Identifying $m=z/w$, we find a complete agreement of \eqref{U1adj} and \eqref{n1inst} up to $m$ independent factor.

	
	{\vskip 20mm}
	\begin{center}
		{\bf Acknowledgements}
	\end{center}
	We would like to thank H.~Awata, T.~Kimura, A.~Mironov, A.~Morozov, H.~Nakajima, Y.~Ohkubo, 
	Y.~Yoshida and Y.~Zenkevich for useful discussions. 
	In particular, we are grateful to H.~Konno and J.~Shiraishi for sharing their insights on the elliptic algebra. 
	Our work is supported in part by Grants-in-Aid for Scientific Research (\# 18K03274) (H.K.).
	The work of P.C. is supported by the MEXT Scholarship. 
	
	\bigskip   

	\titleformat{\section}[block]{\large\scshape\centering}{\appendixname~\thesection .}{0.5em}{}
	[\HRule{3pt}] 
	\appendix

	\section{ Normalization of the intertwiner  
	}
	\label{appendixA}
	
	In this appendix, we provide a detailed computation of the normalization 
	factor $ \cals{G}_{\lambda} $ of the intertwiner. To set the stage, we first fix notations and provide some definitions.
	For a non-negative integer $ m $ we define
	\begin{ceqn}
		\begin{gather}
			B_m(v)
			:= 
			\exp\bigg(
			-\sum_{n=1}^{\infty}\frac{1}{n}\frac{1-p^n}{1-p^n_*}q_2^{nm}\gq^{n/2}\widetilde{a}_{-n}v^n
			\bigg)
			\exp\bigg(
			\sum_{n=1}^{\infty}\frac{1}{n}\widetilde{a}_nv^{-n}q_2^{-nm}\gq^{-3n/2}
			\bigg).
			\label{A1}
		\end{gather}
	\end{ceqn}
	For a partition $ \lambda $, and for $ n \geq \ell(\lambda) $ we define
	\begin{gather}
		\widetilde{\bb{I}}^{[n]}_{\lambda}(v)
		:=
		\widetilde{\bb{I}}_{\lambda_1 - 1}(v;p)\widetilde{\bb{I}}_{\lambda_2 - 1}(q_2v;p)\cdots \widetilde{\bb{I}}_{\lambda_{n} - 1}(q_2^{n  - 1}v;p),
		\label{A2}
	\end{gather}
	where
	\begin{align}
		\widetilde{\bb{I}}_m(z;p) =~\normord{
			\prod_{l = 1}^{\infty}\eta(q_1^{m+ l}z;p)}
	\end{align}
	and $ \eta(z;p) $ is the oscillator part  of the horizontal representation of $E(z;p)$ (see \eqref{3.38mainreport}). 
	Up to the zero mode factor, $\widetilde{\bb{I}}_m(z;p)$ gives a component of the intertwiner 
	for the vector representation \cite{Awata:2018svb}, \cite{Cheewaphutthisakun:2021cud}. 
	From the definition it is straightforward to show that for $ n \geq \ell(\lambda) $
	\begin{gather}
		\normord{
			\widetilde{\bb{I}}^{[n]}_{\lambda}(v)B_n(v)
		}
		= 
		\normord{
			\widetilde{\bb{I}}^{[n+1]}_{\lambda}(v)B_{n+1}(v)
		}.
		\label{a4mainpaper}
	\end{gather}
	\begin{rem}
		We can rewrite the intertwiner \eqref{4.19mainpaper} as
		\begin{ceqn}
			\begin{align}
				\Psi_{\lambda}(v;p) 
				=& z_{\lambda}\cals{G}^{-1}_{\lambda}
				\normord{
					\widetilde{\bb{I}}^{[\ell(\lambda)]}_{\lambda}(v)
					B_{\ell(\lambda)}(v)
				}
				\notag \\ 
				=& z_{\lambda}\cals{G}^{-1}_{\lambda}
				\normord{\,
					\widetilde{\bb{I}}_{\lambda_1 - 1}(v;p)\widetilde{\bb{I}}_{\lambda_2 - 1}(q_2v;p)\cdots 
					\widetilde{\bb{I}}_{\lambda_{\ell(\lambda)} - 1}(q_2^{\ell(\lambda) - 1}v;p)B_{\ell(\lambda)}(v)\,
				}. 
				\label{1.2workbook}
			\end{align}
		\end{ceqn}
		Then, from \eqref{a4mainpaper} we get that
		\begin{ceqn}
			\begin{gather}
				\Psi_{\lambda}(v;p) = z_{\lambda}\cals{G}^{-1}_{\lambda}
				\normord{
					\widetilde{\bb{I}}^{[n]}_{\lambda}(v)B_n(v)
				}
				\quad (n \geq \ell(\lambda)). 
			\end{gather}
		\end{ceqn}
	\end{rem}
	
	\medskip 
	
	Now following \cite{Awata:2018svb}, we define $ \cals{G}_{\lambda} $ 
	to be the coefficient appearing by removing the normal ordering of $ \normord{
		\widetilde{\bb{I}}^{[n]}_{\lambda}(v)B_n(v)
	} $. More precisely, for $ n > \ell(\lambda) $, 
	\begin{align}
		\normord{
			\widetilde{\bb{I}}^{[n]}_{\lambda}(v)B_n(v)
		}
		&\defeq
		\cals{G}^{[n]}\cals{G}_{\lambda}
		\widetilde{\bb{I}}^{[n]}_{\lambda}(v)B_n(v),
		\label{eqn1.9}
	\end{align}
	where $ \cals{G}^{[n]} $ is defined by 
	\begin{align}
		\normord{
			\widetilde{\bb{I}}^{[n]}_{\emptyset}(v)B_n(v)
		}
		&\defeq 
		\cals{G}^{[n]}
		\widetilde{\bb{I}}^{[n]}_{\emptyset}(v)B_n(v). 
		\label{1.8}
	\end{align}
	Namely we factorize the coefficient into the $\lambda$ dependent part $\cals{G}_{\lambda}$ 
	and the $n$ dependent part $\cals{G}^{[n]} $. 
	Hence, $ \cals{G}_{\lambda} $ should be independent of $ n $ as long as $ n > \ell(\lambda) $. 
	
	From \eqref{A1} and \eqref{A2}, it is straightforward to show that for $ m > \ell(\lambda) $, 
	\begin{equation}
		\widetilde{\bb{I}}_{\emptyset}^{[m]}(v;p)B_{m}(v)
		=
		\exp\bigg(
		-\sum_{n=1}^{\infty}\frac{1}{1-q_1^{-n}}\frac{1}{n}\frac{1-p^n}{1-p^n_*}q_2^nm
		\bigg)
		\normord{
			\widetilde{\bb{I}}_{\emptyset}^{[m]}(v;p)B_{m}(v)
		}.
	\end{equation}
	Thus, we obtain that 
	\begin{align}
		\cals{G}^{[m]} = \exp\bigg(
		\sum_{n=1}^{\infty}\frac{1}{1-q_1^{-n}}\frac{1}{n}\frac{1-p^n}{1-p^n_*}q_2^nm
		\bigg). 
	\end{align}
	Similarly,  replacing $\widetilde{\bb{I}}_{\emptyset}^{[m]}(v;p)$ by $\widetilde{\bb{I}}_{\lambda}^{[m]}(v;p)$
	we can show 
	\begin{align}\label{1.26}
		\cals{G}^{[m]}\cals{G}_{\lambda}
		=& \prod_{k=1}^{m-1}
		\exp\bigg(
		-\sum_{n=1}^{\infty}\sum_{l=0}^{k-1}\frac{1}{n}\frac{q_2^n - 1}{1-q_1^{-n}}
		\frac{1-p^n}{1-p^n_*}q_2^{n(k-l)}q_1^{n(\lambda_{k+1} - \lambda_{l+1})}
		\bigg)
		\notag \\
		&\cdot \prod_{k=1}^{m}
		\exp\bigg(
		\sum_{n=1}^{\infty}\frac{1}{n}\frac{1-p^n}{1-p^n_*}q^{nm}_2\frac{1}{1-q_1^{-n}}(q_1^{\lambda_k}q_2^{k-1})^{-n}
		\bigg).
	\end{align}
	According to (2) in page 338 of \cite{MacD}, we have the following identity:
	\begin{align}
		(1-q)\sum_{s \in \lambda}q^{a(s)}t^{\ell(s) + 1}
		= \sum_{i=1}^{n}(t - q^{\lambda_i}t^{n+1 - i}) - (1-t)\sum_{1 \leq i < j \leq n}q^{\lambda_i - \lambda_j}t^{j-i},
	\end{align}
	where $ a(s)$ and $\ell(s) $ are the arm-length and the leg-length of the box $ s $ in the partition $ \lambda $, respectively. Equivalently, 
	\begin{ceqn}
		\begin{align}
			&\sum_{k=1}^{m-1}\sum_{l=0}^{k-1}q_2^{n(k-l)}q_1^{n(\lambda_{k+1} - \lambda_{l + 1})}
			\notag \\
			& = \frac{1}{1-q_2^n}
			\bigg[
			\sum_{i=1}^{m}(q_2^n - q_1^{-n\lambda_i}q_2^{n(m+1 - i)})
			- (1-q_1^{-n})\sum_{s\in\lambda}q_1^{-na(s)}q_2^{n(\ell(s) + 1)}
			\bigg].
			\label{1.32}
		\end{align}
	\end{ceqn}
	Applying \eqref{1.32} to \eqref{1.26}, we obtain that
	\begin{ceqn}
		\begin{align}
			&\cals{G}^{[m]}\cals{G}_{\lambda}
			\notag \\ 
			&=
			\exp\bigg(
			\sum_{n=1}^{\infty}\frac{1}{n}\frac{1}{1-q_1^{-n}}\frac{1-p^n}{1-p^n_*}
			\bigg[
			\sum_{i=1}^{m}(q_2^n - q_1^{-n\lambda_i}q_2^{n(m+1 - i)})
			- (1-q_1^{-n})\sum_{s\in\lambda}q_1^{-na(s)}q_2^{n(\ell(s) + 1)}
			\bigg]
			\bigg)
			\notag \\ 
			&\quad \cdot \prod_{k=1}^{m}
			\exp\bigg(
			\sum_{n=1}^{\infty}\frac{1}{n}\frac{1-p^n}{1-p^n_*}q^{nm}_2\frac{1}{1-q_1^{-n}}(q_1^{\lambda_k}q_2^{k-1})^{-n}
			\bigg).
			\notag \\ 
			&= 
			\exp\bigg(
			\sum_{n=1}^{\infty}\frac{1}{n}\frac{1}{1-q_1^{-n}}\frac{1-p^n}{1-p^n_*}
			\bigg[
			\sum_{i=1}^{m}(q_2^n - q_1^{-n\lambda_i}q_2^{n(m+1 - i)})
			- (1-q_1^{-n})\sum_{s\in\lambda}q_1^{-na(s)}q_2^{n(\ell(s) + 1)}
			\bigg]
			\bigg)
			\notag \\ 
			&\quad \cdot \exp\bigg(
			\sum_{n=1}^{\infty}\sum_{k=1}^{m}\frac{1}{n}\frac{1-p^n}{1-p^n_*}\frac{1}{1-q_1^{-n}}q_2^{n(m-k+1)}q_1^{-n\lambda_k}
			\bigg)
			\notag \\ 
			&= 
			\exp\bigg(
			\sum_{n=1}^{\infty}\frac{1}{n}\frac{1}{1-q_1^{-n}}\frac{1-p^n}{1-p^n_*}
			\bigg[
			q_2^nm
			- (1-q_1^{-n})\sum_{s\in\lambda}q_1^{-na(s)}q_2^{n(\ell(s) + 1)}
			\bigg]
			\bigg).
		\end{align}
	\end{ceqn}
	So we get that 
	\begin{align}
		\cals{G}_{\lambda} 
		= 
		\exp\bigg(
		-\sum_{n=1}^{\infty}\frac{1}{n}\frac{1-p^n}{1-p^n_*}
		\bigg[
		\sum_{s\in\lambda}q_1^{-na(s)}q_2^{n(\ell(s) + 1)}
		\bigg]
		\bigg).
	\end{align}
	From this we see that $ \cals{G}_{\lambda} $ does not depend on $ m $, as should be.

	\section{Intertwining relations for $E$ and $F$}

	The free boson oscillator part of the horizontal representation is 
	\begin{align}
		\eta(z;p)
		=
		&
		\exp\bigg(
		\sum_{n=1}^{\infty}\frac{\kappa_n}{n}\frac{\gq^{-n/2}(1-p^n)}{(1-p^n_*)(\gq^n - \gq^{-n})}\widetilde{a}_{-n}z^n
		\bigg)\cdot\exp\bigg(
		-\sum_{n=1}^{\infty}\frac{\kappa_n}{n}\frac{\gq^{-n/2}}{\gq^n - \gq^{-n}}\widetilde{a}_{n}z^{-n}
		\bigg),
		\\
		\xi(z;p) 
		=
		&
		\exp\bigg(
		-\sum_{n=1}^{\infty}\frac{\kappa_n}{n}\frac{\gq^{n/2}}{\gq^n - \gq^{-n}}\widetilde{a}_{-n}z^n
		\bigg) \cdot \exp\bigg(
		\sum_{n=1}^{\infty}\frac{\kappa_n}{n}\frac{\gq^{n/2}(1-p^n_*)}{(1-p^n)(\gq^n - \gq^{-n})}\widetilde{a}_nz^{-n}
		\bigg).
	\end{align}
	With the standard choice of the zero modes the elliptic currents are given by 
	\begin{equation}
		E(z;p) = \frac{(\gq/z)^N u}{(1- q_1)(1- q_2)} \eta(z;p), \qquad F(z;p) = \frac{(\gq/z)^{-N} u^{-1}}{(1- q_1^{-1})(1- q_2^{-1})}\xi(z;p),
	\end{equation}
	where they act on the horizontal Fock space with level $N$ and the spectral parameter $u$. 
	Note that the zero modes are independent of the elliptic parameter $p$. 
	We have the following OPE relations with the intertwiner;
	\begin{align}
		\eta(z;p) \Psi_\lambda (v;p) 
		&= \exp \left( \sum_{n=1}^\infty \frac{1}{n} \left( \frac{v}{z} \right)^n 
		\frac{(1- p^n)(1- q_2^n)}{1-p_*^n} \sum_{i=1}^\infty q_1^{n\lambda_i} q_2^{n(i-1)}
		\right)
		\notag \\
		&~~~ :\eta(z;p) \Psi_\lambda (v;p):,
		\\
		\Psi_\lambda (v;p) \eta(z;p) 
		&=	\exp \left(- \sum_{n=1}^\infty \frac{1}{n} \left( \frac{z}{v} \right)^n 
		\frac{(1- p^n)(1- q_2^n)}{1-p_*^n} \sum_{i=1}^\infty q_1^{-n(\lambda_i-1)} q_2^{-n(i-1)}
		\right)
		\notag \\
		&~~~:\eta(z;p) \Psi_\lambda (v;p):,
		\\
		\xi(z;p) \Psi_\lambda (v;p) 
		&= \exp \left( -\sum_{n=1}^\infty \frac{1}{n} \left( \frac{v}{z} \right)^n 
		\gq^n (1- q_2^n)\sum_{i=1}^\infty q_1^{n\lambda_i} q_2^{n(i-1)}
		\right)
		\notag \\
		&~~~ :\xi(z;p) \Psi_\lambda (v;p):,
		\\
		\Psi_\lambda (v;p) \xi(z;p) 
		&= \exp \left( \sum_{n=1}^\infty \frac{1}{n} \left( \frac{z}{v}\right)^n 
		\gq^n (1- q_2^n)\sum_{i=1}^\infty q_1^{-n(\lambda_i -1)} q_2^{-n(i-1)}
		\right)
		\notag \\
		&~~~ :\xi(z;p) \Psi_\lambda (v;p):.
	\end{align}

	Assume that $\Psi_{\lambda}(v;p)$ maps level $N$ and the spectral parameter $u$
	to $N+1$ and $w$. The intertwining relation for the vacuum component $F(z;p) \Psi_\varnothing(v;p) 
	= \Psi_\varnothing(v;p) F(z;p)$ implies
	\begin{equation}
		( \gq/z)^{-N-1} w^{-1} \left( 1 - \frac{\gq v}{z} \right) = ( \gq/z)^{-N} u^{-1} \left( 1 - \frac{z}{\gq v} \right).
	\end{equation}
	Hence we obtain the condition $w= -uv$ for the existence of $\Psi_{\lambda}(v;p)$.
	
	Now let us turn to the intertwining relation for $E(z;p)$;
	\begin{align}
		E(z;p)\Psi_{\lambda}(v;p) &= 
		\sum_{k=1}^{\ell(\lambda) + 1}\langle\lambda+1_k|E(z;p_*)|\lambda\rangle \Psi_{\lambda+1_k}(v;p) 
		\notag \\
		&\qquad +\langle \lambda|K^-(z;p_*)|\lambda\rangle \Psi_{\lambda}(v;p)E(z;p),
	\end{align}
	Up to the zero mode factors we have
	\begin{align}
		&\eta(z;p) \Psi_\lambda (v;p) 
		\notag \\
		&= \prod_{i=1}^\infty \frac{(p_* q_1^{\lambda_i-1} q_2^{i-2} (v/z);p_*)_\infty(q_1^{\lambda_i} q_2^{i} (v/z);p_*)_\infty }
		{(q_1^{\lambda_i} q_2^{i-1} (v/z);p_*)_\infty(p_* q_1^{\lambda_i-1} q_2^{i-1} (v/z);p_*)_\infty }
		:\eta(z;p) \Psi_\lambda (v;p):,
	\end{align}
	and
	\begin{align}
		& \langle \lambda|K^-(z;p_*)|\lambda\rangle \Psi_{\lambda}(v;p) \eta(z;p) 
		\notag \\
		&= \gq \prod_{i=1}^\infty \frac{\theta_{p_*} (q_1^{-\lambda_i} q_2^{-i}(z/v)) \theta_{p_*} (q_1^{-\lambda_i+1} q_2^{-i+2}(z/v)) }
		{\theta_{p_*} (q_1^{-\lambda_i} q_2^{-i+1}(z/v)) \theta_{p_*} (q_1^{-\lambda_i+1} q_2^{-i+1}(z/v)) }
		\notag \\
		&\prod_{i=1}^\infty \frac{(q_1^{-\lambda_i+1} q_2^{-i+1} (z/v);p_*)_\infty (p_* q_1^{-\lambda_i} q_2^{-i+1} (z/v);p_*)_\infty }
		{(q_1^{-\lambda_i+1} q_2^{-i+2} (z/v);p_*)_\infty (p_* q_1^{-\lambda_i} q_2^{-i} (z/v);p_*)_\infty }
		:\eta(z;p) \Psi_\lambda (v;p):
		\notag \\
		&= \gq \prod_{i=1}^\infty \frac{(1-q_1^{-\lambda_i} q_2^{-i} (z/v))} {(1-q_1^{-\lambda_i} q_2^{-i+1} (z/v))}
		\frac{(p_* q_1^{\lambda_i} q_2^{i}(v/z);p_*)_\infty (p_* q_1^{\lambda_i-1} q_2^{i-2}(v/z);p_*)_\infty}
		{(p_* q_1^{\lambda_i} q_2^{i-1}(v/z);p_*)_\infty (p_* q_1^{\lambda_i-1} q_2^{i-1}(v/z);p_*)_\infty}
		\notag \\
		&\qquad :\eta(z;p) \Psi_\lambda (v;p):.
	\end{align}
	Hence, incorporating the zero mode factor with $w=-uv$, we find
	\begin{align}\label{comE}
		& (1-q_1) (1-q_2) \left( E(z;p) \Psi_\lambda (v;p) - \langle \lambda|K^-(z;p_*)|\lambda\rangle \Psi_{\lambda}(v;p) E(z;p) \right)
		\notag \\
		&= \left( \frac{\gq}{z} \right)^{N+1} (-uv) \prod_{i=1}^\infty
		\frac{(p_* q_1^{\lambda_i} q_2^{i}(v/z);p_*)_\infty (p_* q_1^{\lambda_i-1} q_2^{i-2}(v/z);p_*)_\infty}
		{(p_* q_1^{\lambda_i} q_2^{i-1}(v/z);p_*)_\infty (p_* q_1^{\lambda_i-1} q_2^{i-1}(v/z);p_*)_\infty}  :\eta(z;p) \Psi_\lambda (v;p):
		\notag \\
		&~\left( \frac{1}{1 - q_2^{\ell(\lambda)}(v/z)} \prod_{i=1}^{\ell(\lambda)} 
		\frac{1- q_1^{\lambda_i} q_2^i(v/z)}{1- q_1^{\lambda_i} q_2^{i-1}(v/z)} 
		-  \frac{-z/v}{1 - q_2^{-\ell(\lambda)}(z/v)} \prod_{i=1}^{\ell(\lambda)} 
		\frac{1- q_1^{-\lambda_i} q_2^{-i} (z/v)}{1- q_1^{-\lambda_i} q_2^{-i+1}(z/v)} \right)~.
	\end{align}
	Note that the last factor in the big parentheses is independent of the elliptic parameter $p$ and 
	$p$ dependence appears only through the factor $(p_* X ; p_*)_\infty$, which  becomes trivial when $p=0$. 
	The overall monomial factor is nothing but the zero mode \eqref{zeromodes} for the {\it target} Fock space. 
	
	Now to evaluate the last factor we use the following lemma; 
	for a rational function of the form
	\beq
	f_{+}(z) = \prod_{i=1}^n \frac{1- \alpha_i z}{1- \beta_i z},
	\eeq
	we assume that $\beta_j$ are all distinct and 
	\beq\label{balance}
	\prod_{i=1}^n \alpha_i = \prod_{i=1}^n \beta_i.
	\eeq
	Namely all the poles of $f_{+}(z)$ are simple and 
	\beq\label{infinity}
	\lim_{z \to \infty} f_{+}(z) =1.
	\eeq
	If a given rational function is
	\beq
	\tilde{f}_{+}(z) = \frac{\prod_{i=1}^m (1- \alpha_i z)}{\prod_{j=1}^{n} (1- \beta_j z)}, \qquad
	\prod_{i=1}^m \alpha_i = \prod_{j=1}^n \beta_j,
	\eeq
	with $m<n$ like \eqref{comE}. We can multiply $(1-z)^{n-m}$ and consider
	\beq
	f_{+}(z) = (z-1)^{n-m}\tilde{f}_{+}(z).
	\eeq
	By using the condition \eqref{balance}, we can see
	\beq
	f_{-}(z) = \prod_{i=1}^n \frac{1- \alpha_i^{-1} z^{-1}}{1- \beta_i^{-1} z^{-1}},
	\eeq
	satisfies
	\beq\label{equivalence}
	f_{+}(z) = f_{-}(z)
	\eeq
	for $z \neq \beta_j^{-1}$. Then the following formula holds;
	\beq\label{decompose}
	f_{+} (z) - f_{-} (z) = \sum_{k=1}^n \delta(\beta_k z) 
	\frac{\prod_{i=1}^n (1- \beta_k^{-1}\alpha_i)}{\prod_{j \neq k} (1- \beta_k^{-1}\beta_j)}.
	\eeq
	Note that the coefficients of the delta function are the residues of $f_{+}(z)$ at the corresponding poles. 

	We can prove \eqref{decompose} as follows;
	since $f_{+}(z)$ is holomorphic on $\mathbb{P}^1$ with only simple poles at $z = \beta_j^{-1}$, the partial
	fraction decomposition of $f_+(z)$ is\footnote{Note \eqref{infinity}.}
	\beq
	f_{+}(z) = 1 + \sum_{k=1}^n \frac{c_k}{1- \beta_k z},
	\qquad \sum_{k=1}^n c_k =0,
	\eeq
	where $c_k$ is the residue of $f_{+}(z)$ at $z = \beta_k^{-1}$.
	Then by \eqref{equivalence}, we must have
	\beq
	f_{-}(z) = 1 -  \sum_{i=1}^n \frac{c_k \beta_k^{-1} z^{-1}}{1- \beta_k^{-1} z^{-1}},
	\eeq
	and hence
	\beq
	f_{+}(z) - f_{-}(z) 
	= \sum_{k=1}^n c_k \left(\frac{1}{1- \beta_k z} + \frac{\beta_k^{-1} z^{-1}}{1- \beta_k^{-1} z^{-1}} \right)
	= \sum_{k=1}^n c_k \delta(\beta_k z).
	\eeq

	We can check that the last factor in \eqref{comE} satisfies the above assumptions. Applying the lemma
	we obtain
	\begin{align}
		& (1-q_1)(1-q_2) \left( E(z;p) \Psi_\lambda (v;p) - \langle \lambda|K^-(z;p_*)|\lambda\rangle \Psi_{\lambda}(v;p) E(z;p) \right)
		\notag \\
		&= \left( \frac{\gq}{q_1^{\lambda_k} q_2^{k-1}v} \right)^{N+1} (-uv) \prod_{i=1}^\infty
		\frac{(p_* q_1^{\lambda_i - \lambda_k} q_2^{i-k+1} ;p_*)_\infty (p_* q_1^{\lambda_i - \lambda_k -1} q_2^{i-k-1} ;p_*)_\infty}
		{(p_* q_1^{\lambda_i -\lambda_k} q_2^{i-k} ;p_*)_\infty (p_* q_1^{\lambda_i - \lambda_k -1} q_2^{i-k} ;p_*)_\infty}
		\notag \\
		& \times (1-q_2) \sum_{k=1}^{\ell(\lambda)+1} \delta\left(q_1^{\lambda_k} q_2^{k-1}\frac{v}{z} \right) 
		\prod_{i \neq k} \frac{1- q_1^{\lambda_i -\lambda_k}q_2^{i-k+1}}{1- q_1^{\lambda_i -\lambda_k}q_2^{i-k}}
		:\eta(q_1^{\lambda_k} q_2^{k-1}v ;p) \Psi_\lambda (v;p):~.
	\end{align}
	Since
	\begin{equation}
		:\eta(q_1^{\lambda_k} q_2^{k-1}v ;p) \Psi_\lambda (v;p):
		= \frac{z_\lambda}{\mathcal{G}_\lambda} \frac{\mathcal{G}_{\lambda + 1_k}} {z_{\lambda+ 1_k}}
		\Psi_{\lambda + 1_k}  (v;p),
	\end{equation} 
	using \eqref{verticalE} and assuming $\langle \lambda + 1_k	|\lambda + 1_k\rangle =1$, 
	we obtain recursion relations for the normalization factor $\mathcal{G}_\lambda$
	and the zero mode factor $z_\lambda$;
	\begin{equation}
		\frac{z_{\lambda + 1_k}} {z_\lambda} = q_2^{k-1} (-uv) \left( \frac{\gq}{q_1^{\lambda_k} q_2^{k-1} v} \right)^{N+1},
	\end{equation}
	and 
	\begin{align}
		&\frac{\mathcal{G}_{\lambda + 1_k}}{\mathcal{G}_\lambda} 
		\prod_{i \neq k}
		\frac{(q_1^{\lambda_i - \lambda_k} q_2^{i-k+1} ;p_*)_\infty (p_* q_1^{\lambda_i - \lambda_k -1} q_2^{i-k-1} ;p_*)_\infty}
		{(q_1^{\lambda_i -\lambda_k} q_2^{i-k} ;p_*)_\infty (p_* q_1^{\lambda_i - \lambda_k -1} q_2^{i-k} ;p_*)_\infty}
		\notag \\
		&\qquad = q_2^{k-1} 
		\prod_{i=1}^{k-1}\frac{\theta_{p_*}(q_1^{\lambda_k - \lambda_i}q_2^{k-i-1})\theta_{p_*}(q_1^{\lambda_k - \lambda_i +1}q_2^{k-i+1})}
		{\theta_{p_*}(q_1^{\lambda_k - \lambda_i}q_2^{k-i})\theta_{p_*}(q_1^{\lambda_k - \lambda_i + 1}q_2^{k-i})}.
	\end{align}
	Note that we have adjusted the monomial factor $q_2^{k-1}$ between $z_\lambda$ and $\mathcal{G}_\lambda$ to simplify the 
	recursion relation for $\mathcal{G}_\lambda$ as follows;
	\begin{align}
		\frac{\mathcal{G}_{\lambda + 1_k}}{\mathcal{G}_\lambda} 
		&= 
		\prod_{i=1}^{k-1}\frac{(p_* q_1^{\lambda_k - \lambda_i}q_2^{k-i-1}; p_*)_\infty (q_1^{\lambda_k - \lambda_i +1}q_2^{k-i+1}; p_*)_\infty}
		{(p_* q_1^{\lambda_k - \lambda_i}q_2^{k-i}; p_*)_\infty (q_1^{\lambda_k - \lambda_i + 1}q_2^{k-i}; p_*)_\infty}
		\notag \\
		& \qquad
		\prod_{i=k+1}^{\infty}
		\frac{(q_1^{\lambda_i -\lambda_k} q_2^{i-k} ;p_*)_\infty (p_* q_1^{\lambda_i - \lambda_k -1} q_2^{i-k} ;p_*)_\infty}
		{(q_1^{\lambda_i - \lambda_k} q_2^{i-k+1} ;p_*)_\infty (p_* q_1^{\lambda_i - \lambda_k -1} q_2^{i-k-1} ;p_*)_\infty}.
	\end{align}
	We note the $p$ dependent factor in the recursion relation is 
	\begin{align}
		&\prod_{i=1}^{k-1}\frac{(p_* q_1^{\lambda_k - \lambda_i}q_2^{k-i-1}; p_*)_\infty (p_* q_1^{\lambda_k - \lambda_i +1}q_2^{k-i+1}; p_*)_\infty}
		{(p_* q_1^{\lambda_k - \lambda_i}q_2^{k-i}; p_*)_\infty (p_* q_1^{\lambda_k - \lambda_i + 1}q_2^{k-i}; p_*)_\infty}
		\notag \\
		&
		\prod_{i=k+1}^{\infty}
		\frac{(p_* q_1^{\lambda_i -\lambda_k} q_2^{i-k} ;p_*)_\infty (p_* q_1^{\lambda_i - \lambda_k -1} q_2^{i-k} ;p_*)_\infty}
		{(p_* q_1^{\lambda_i - \lambda_k} q_2^{i-k+1} ;p_*)_\infty (p_* q_1^{\lambda_i - \lambda_k -1} q_2^{i-k-1} ;p_*)_\infty}.
	\end{align}
	It is striking that this is the same as the remainder factor $R_\lambda^{(k)}$ appearing in section \ref{subsect3.3}
	by the change of variables $(q_1, q_2, p) \to (q_1^{-1}, q_2^{-1}, p_*)$. This means the base change discussed there 
	eliminates the above $p$ dependence by employing the "second" formula for the vertical Fock representation derived in section \ref{subsect3.3}. 
	When $p \to 0$ or after the base change, the recursion reduces to
	\begin{align}
		&\frac{\mathcal{G}_{\lambda + 1_k}}{\mathcal{G}_\lambda} 
		= \prod_{i=1}^{k-1}\frac{1- q_1^{\lambda_k - \lambda_i +1}q_2^{k-i+1}}
		{1- q_1^{\lambda_k - \lambda_i + 1}q_2^{k-i}}
		\prod_{i=k+1}^{\infty}\frac{1- q_1^{\lambda_i -\lambda_k} q_2^{i-k}}
		{1- q_1^{\lambda_i - \lambda_k} q_2^{i-k+1}}.
	\end{align}
	With the initial condition $\mathcal{G}_\varnothing =1$ the recursion relation is solved by
	\begin{equation}
		\mathcal{G}_\lambda (q_1, q_2)  = \prod_{s \in \lambda} ( 1- q_1^{-a(s)} q_2^{\ell(s) +1}).
	\end{equation}
	After $(q_1, q_2) \to (q_1^{-1}, q_2^{-1})$, it agrees with the standard normalization factor 
	for the integral form of the Macdonald function. 
	In general for $p \neq 0$ the solution of the recursion relation is 
	\begin{equation}
		\mathcal{G}_\lambda (q_1, q_2 ; p )  = \frac{\prod_{s \in \lambda} (q_1^{-a(s)} q_2^{\ell(s) +1}; p_*)_\infty}
		{\prod_{s \in \lambda} (p q_1^{-a(s)} q_2^{\ell(s) +1}; p_*)_\infty},
	\end{equation}
	which agrees with the result in Appendix A. 
	Finally the recursion relation for $z_\lambda$ is solved by
	\begin{equation}
		z_\lambda = q_2^{n(\lambda)} (-v)^{-N \vert \lambda \vert} u^{\vert \lambda \vert} f_\lambda(q_1, q_2)^{-N-1},
	\end{equation}
	where 
	\begin{equation}
		n(\lambda) := \sum_{k=1}^\infty  \lambda_k (k-1). 
	\end{equation}


	\section{Free field representation and $SU(4)$ Omega background}
	\label{SU4}
	
	In this appendix we show an interesting connection of the free field representation 
	employed in this paper and $SU(4)$ Omega background $(q_1,q_2,q_3,q_4)$ or 
	the equivariant parameters of the toric action on $\mathbb{C}^4$,
	which opens a way of interpreting our computation from the viewpoint of 
	eight dimensional gauge theory, or the spiked instantons of Nekrasov \cite{Nekrasov:2015wsu}. 	
	To motivate the $SU(4)$ Omega background,
	let us look at the affine quiver $\hat{A}_0$ with a single node and a single loop (a.k.a.
	the Jordan quiver). Since it has a single node, the $\gq$-deformed Cartan matrix has a single 
	component $C= (1- \mu^{-1})(1- \gq^{-1} \mu)$, where $\mu$ is a parameter associated with the
	loop of $\hat{A}_0$ quiver. 
	The quiver gauge theory for $\hat{A}_0$ is the supersymmetric gauge theory with adjoint hypermultiplet,
	usually called $\mathcal{N}=2^{*}$ theory and the parameter $\mu$ is physically the exponentiated mass
	parameter. 
	If we follow the prescription of \cite{FRW} and \cite{Kimura:2015rgi}, the commutation relation of
	the so-called \lq\lq root boson\rq\rq\ takes the form;
	\beqa
	\left [ \lambda_n. \lambda_m \right] &=& -n (1-q_1)(1-q_2) C \delta_{n+m, 0} \CR
	&=& -n (1-q_1)(1-q_2) (1-q_3)(1-q_4)\delta_{n+m, 0},
	\eeqa
	where we have defined $q_3= \mu^{-1}$ and $q_4 = \gq^{-1} \mu$ with $\gq = q_1 q_2$. 
	We are going to show that for each pair $(ij)$ with $1 \leq i <  j \leq 4$, there exists 
	a Fock representation of the quasi-Hopf twist of DIM algebra with the central charge $C= \sqrt{q_i q_j}$. 
	Thus, we obtain a family of six Fock representations with various central charges, which
	seems to match with the six stacks of $D3$ configuration for the spiked 
	instanton, where the pair $(ij)$ specifies a choice of codimension four subspace of
	$\mathbb{C}^4$ which are fixed by the toric action 
	\beq
	(z_1, z_2, z_3, z_4) \longrightarrow (q_1z_1, q_2z_2, q_3z_3, q_4z_4), \qquad q_1 q_2 q_3 q_4 =1.
	\eeq
	
	One can define a completely $S_4$ symmetric deformed Heisenberg algebra
	\beq
	\left[ a_n, a_m \right] = - n (1-q_1^{\pm n})(1-q_2^{\pm n})(1-q_3^{\pm n})(1-q_4^{\pm n})\delta_{n+m,0}, 
	\eeq
	and an $S_4$ symmetric vertex operator 
	\beq\label{mother}
	\Phi(z) =  \exp \left( \sum_{n=1}^\infty \frac{a_{-n}}{n} z^{n} \right)  
	\exp \left( \sum_{n=1}^\infty \frac{a_{n}}{n} z^{-n} \right).
	\eeq
	Then we define a quartet of the screening operators by
	\beq
	S^{(k)} (z) =  \exp \left( \sum_{n=1}^\infty \frac{a_{-n}}{n(1- q_k^{n})} z^{n} \right) 
	\exp \left( \sum_{n=1}^\infty \frac{a_{n}}{n(1- q_k^{-n}) } z^{-n} \right), \qquad (1 \leq k \leq 4)
	\eeq
	which satisfies 
	\beq
	\Phi(z) = : S^{(k)}(z) S^{(k)} (q_k z)^{-1}:.
	\eeq
	Writing the screening operator as 
	\beq
	S^{(k)} (z) =  \exp \left( \sum_{n=1}^\infty \frac{1}{n} s^{(k)}_{-n} z^{n} \right) 
	\exp \left( \sum_{n=1}^\infty \frac{1}{n} s^{(k)}_{n} z^{-n} \right),
	\eeq
	we have the commutation relation
	\beq
	\left[ s^{(k)}_n, s^{(k)}_m \right] = - n \frac{\prod_{i \neq k} (1- q_i^{\mp n})}
	{1- q_k^{\pm n}}\delta_{n+m,0}.
	\eeq
	Computing OPE coefficients we obtain
	\beq\label{SkSk}
	S^{(k)}(z)S^{(k)}(w) 
	= \frac{\theta_{q_k}(w/z)\theta_{q_k}(q_i^{-1}q_j^{-1}w/z) \theta_{q_k}(q_i^{-1}z/w) \theta_{q_k}(q_j^{-1}z/w) }
	{\theta_{q_k}(z/w)\theta_{q_k}(q_i^{-1}q_j^{-1}z/w)\theta_{q_k}(q_i^{-1}w/z) \theta_{q_k}(q_j^{-1}w/z) } S^{(k)}(w)S^{(k)}(z),
	\eeq 
	where $\{ i,j, k,\ell \}= \{ 1,2,3,4 \}$.
	Using the inversion formula \eqref{theta-inversion},  we can rewrite the relation \eqref{SkSk} as
	\beq
	S^{(k)}(z)S^{(k)}(w) = G^{(ij)}(w/z; q_k) S^{(k)}(w)S^{(k)}(z),
	\eeq
	where 
	\beq
	G^{(ij)}(u; q_k):= \frac{\theta_{q_k} (q_i u)\theta_{q_k} (q_j u)\theta_{q_k} (q_i^{-1} q_j^{-1} u)}
	{\theta_{q_k} (q_i^{-1} u)\theta_{q_k} (q_j^{-1} u)\theta_{q_k} (q_i q_j u)}.
	\eeq
	Note that since $\theta_{p^{-1}} (u) = \theta_{p} (u^{-1})^{-1}$, we have
	\beq\label{G-inversion}
	G^{(ij)}(u; q_k^{-1}) =  G^{(ij)}(u; q_k)^{-1}.
	\eeq
	$G^{(ij)}$ is related to the structure function used in this paper by $G^{(12)} = \mathcal{G}^{-1}$. 
	Later we will take $q_3 = p_*^{-1}$ and \eqref{G-inversion} shows the consistency of this choice.

	The commutation relation of the screening operators of different kind generates 
	a new operator, which we identify with the Cartan current.
	For convenience let us denotes the OPE factors of two operators $A(z)$ and $B(w)$ by $c(A(z), B(w))$,
	namely
	\beq
	A(z) B(w) = c(A(z), B(w)) : A(z) B(w) :, \qquad |z| > |w|.
	\eeq
	We have
	\beqa
	c(S^{(k)}(z), S^{(\ell)}(w)) &=& \frac{\left(1- q_k q_i \frac{w}{z}\right)\left(1- q_k q_j \frac{w}{z}\right)}
	{\left(1- q_k \frac{w}{z}\right)\left(1- q_\ell^{-1} \frac{w}{z}\right)}, \CR
	c(S^{(\ell)}(w), S^{(k)}(z)) &=& \frac{\left(1- q_\ell q_i \frac{z}{w}\right)\left(1- q_\ell q_j \frac{z}{w}\right)}
	{\left(1- q_k^{-1} \frac{z}{w}\right)\left(1- q_\ell \frac{z}{w}\right)}.
	\eeqa
	Hence the support of the commutation relation $[S^{(k)}(z), S^{(\ell)}(w)]$ is only at the simple poles;
	$q_i w/z =1$ and $q_j^{-1}w/z=1$. 
	Computing the residues there, we find
	\beqa
	&& \left[ S^{(k)}(z), S^{(\ell)}(w) \right] \CR
	&=& \frac{(1-q_i)(1-q_j)}{1-q_i q_j} 
	\left( \frac{1}{1- q_k \frac{w}{z}} - \frac{1}{1- q_\ell^{-1} \frac{w}{z}} 
	+ \frac{1}{1- q_k^{-1} \frac{z}{w}} - \frac{1}{1- q_\ell \frac{z}{w}} \right)
	: S^{(k)}(z) S^{(\ell)}(w) : \CR
	&=& \frac{(1-q_i)(1-q_j)}{1-q_i q_j} 
	\left( \delta\left(q_k \frac{w}{z}\right) - \delta\left( q_\ell \frac{z}{w}\right) \right)
	: S^{(k)}(z) S^{(\ell)}(w) :.
	\eeqa
	Hence introducing 
	\beq
	\psi^{(k\ell)}(z) := : S^{(k)}(q_k z) S^{(\ell)}(z): = : S^{(k)}(z) S^{(\ell)}(q_\ell z):,
	\eeq
	we can express the commutation relation as
	\beq
	\left[ S^{(k)}(z), S^{(\ell)}(w) \right] = \frac{(1-q_i)(1-q_j)}{1-q_i q_j} 
	\left( \delta\left(q_k \frac{w}{z}\right) \psi^{(k\ell)}(w) 
	- \delta\left( q_\ell \frac{z}{w}\right) \psi^{(k \ell)}(z) \right).
	\eeq
	More explicitly the Cartan current is\footnote{After the quasi-Hopf twist, we can treat $\psi^{\pm}$ on an equal footing.}
	\beq
	\psi^{(k\ell)}(z) = \exp \left( \sum_{n=1}^\infty \frac{1-q_k^n q_\ell^n}{n(1-q_k^n)(1-q_\ell^n)} a_{-n} z^n \right)
	\exp \left( \sum_{n=1}^\infty \frac{1-q_k^{-n} q_\ell^{-n}}{n(1-q_k^{-n})(1-q_\ell^{-n})} a_{n} z^{-n} \right).
	\eeq
	To express the OPE factor of the Cartan current, it is convenient to {\it temporally} use the notation
	\beq
	\gq_1 = q_i, \quad \gq_2 = q_j, \quad \gq_3 = q_i^{-1} q_j^{-1}.
	\eeq
	Then we have
	\beq
	c( \psi^{(k\ell)}(z), \psi^{(k\ell)}(w)) = \prod_{m=1}^3 
	\frac{\left(\gq_m \frac{q_k w}{z}; q_k, q_\ell^{-1} \right)_\infty \left(\gq_m^{-1} \frac{w}{q_\ell z}; q_k, q_\ell^{-1} \right)_\infty}
	{\left(\gq_m \frac{w}{q_\ell z}; q_k, q_\ell^{-1} \right)_\infty \left(\gq_m^{-1} \frac{q_k w}{z}; q_k, q_\ell^{-1} \right)_\infty}.
	\eeq
	Using the elliptic gamma function,  we can simplify the exchange relation
	\beq
	\psi^{(k\ell)}(z) \psi^{(k\ell)}(w) = \prod_{m=1}^3 
	\frac{\Gamma_{q_k, q_\ell^{-1}}(\gq_m \frac{w}{q_\ell z})
		\Gamma_{q_k, q_\ell^{-1}}(\gq_m^{-1} \frac{q_k w}{z})}
	{\Gamma_{q_k, q_\ell^{-1}}(\gq_m \frac{q_k w}{z}) \Gamma_{q_k, q_\ell^{-1}}(\gq_m^{-1} \frac{w}{q_\ell z})}
	\psi^{(k\ell)}(w) \psi^{(k\ell)}(z).
	\eeq
	Finally by the difference relation, 
	\beq
	\Gamma_{p_1, p_2}(p_1 u) = \theta_{p_2}(u) \Gamma_{p_1, p_2}(u), \qquad
	\Gamma_{p_1, p_2}(p_2 u) = \theta_{p_1}(u) \Gamma_{p_1, p_2}(u)
	\eeq
	we obtain
	\beqa
	\psi^{(k\ell)}(z) \psi^{(k\ell)}(w) &=& \prod_{m=1}^3 
	\frac{G^{(ij)}(\frac{w}{z}; q_k)}{G^{(ij)}(\frac{w}{z}, q_\ell^{-1})} \psi^{(k\ell)}(w) \psi^{(k\ell)}(z) \CR
	&=& G^{(ij)}(\frac{w}{z}; q_k) G^{(ij)}(\frac{w}{z}, q_\ell) \psi^{(k\ell)}(w) \psi^{(k\ell)}(z),
	\eeqa
	which is manifestly symmetric in $i \leftrightarrow j$ and $k \leftrightarrow \ell$. 
	In the present $S_4$ symmetric formulation the exchange relations of $S^{(k)}(z), S^{(\ell)}(z)$ and 
	$\psi^{(k\ell)}(w)$ are\footnote{Note that $\psi^{(k\ell)}= \psi^{(\ell k)}$.}
	\begin{align}
		S^{(k)}(z) \psi^{(k\ell)}(w) &= G^{(ij)}( q_k w/z ; q_k) \psi^{(k\ell)}(w) S^{(k)}(z)
		= G^{(ij)}(w/z ; q_k) \psi^{(k\ell)}(w) S^{(k)}(z), \notag \\
		S^{(\ell)}(z) \psi^{(k\ell)}(w) &= G^{(ij)}( q_\ell w/z ; q_\ell) \psi^{(k\ell)}(w) S^{(\ell)}(z)
		= G^{(ij)}( w/z ; q_\ell) \psi^{(k\ell)}(w) S^{(\ell)}(z),
	\end{align}
	where we have used $G^{(ij)}(p u ; p) = G^{(ij)}(u ; p)$ which follows from $\theta_p(px) = - x^{-1} \theta_p(x)$.

	In summary, we have obtained a {\it sextet} of the Fock representations of the quasi-Hopf twist of DIM algebra;
	$\mathcal{F}^{(k\ell)}=\mathcal{F}^{(\ell k)}~(1 \leq k < \ell \leq 4)$ 
	generated by $S^{(k)}(z), S^{(\ell)}(z)$ and $\psi^{(k\ell)}(z)$.
	Their commutation relations are;
	\begin{align}
		S^{(k)}(z)S^{(k)}(w) &= G^{(ij)}(\frac{w}{z}; q_k) S^{(k)}(w)S^{(k)}(z), \\
		\psi^{(k\ell)}(z) \psi^{(k\ell)}(w) &= G^{(ij)}(\frac{w}{z}; q_k) G^{(ij)}(\frac{w}{z}, q_\ell) \psi^{(k\ell)}(w) \psi^{(k\ell)}(z), \\
		S^{(k)}(z) \psi^{(k\ell)}(w) &= G^{(ij)}(\frac{w}{z} ; q_k) \psi^{(k\ell)}(w) S^{(k)}(z), \\
		\left[ S^{(k)}(z), S^{(\ell)}(w) \right] &= \frac{(1-q_i)(1-q_j)}{1-q_i q_j} 
		\left( \delta\left(q_k \frac{w}{z}\right) \psi^{(k\ell)}(w) - \delta\left( q_\ell \frac{z}{w}\right) \psi^{(k \ell)}(z) \right).
	\end{align}
	where $\{i, j, k, \ell \} = \{ 1,2, 3, 4 \}$. In fact one can check that 
	\beqa
	&&E(z) = S^{(k)} ( \sqrt{q_k} z), \qquad F(z) = S^{(\ell)} ( \sqrt{q_\ell} z), \\
	&& K^{+}(z) = \psi^{(k\ell)}(z/ \sqrt{q_k} ), \qquad K^{-}(z) = \psi^{(k\ell)}(z/  \sqrt{q_\ell} ),
	\eeqa
	gives a representation of the quasi-Hopf twist of DIM algebra with the central charge $C=\sqrt{q_k q_\ell}$
	and the following correspondence of the structure functions;
	\beq
	G^{(ij)}(u; q_k) \longleftrightarrow \mathcal{G}(u; p_*),
	\qquad
	G^{(ij)}(u; q_\ell)  \longleftrightarrow \mathcal{G}(u; p)^{-1}.
	\eeq
	The normalization of the commutation relation $[E(z), F(z)]$ is
	\beq
	\tilde{g} = \frac{(1-q_i)(1-q_j)}{1-q_i q_j}.
	\eeq
	%
	
	For example, take $SU(4)$ parameters;
	\beq
	q_1 = q, \qquad q_2 = t^{-1}, \qquad q_3 = p_*^{-1}, \qquad q_4 = p, 
	\eeq
	then $C= \sqrt{p/p_*} = \gq$ and for $p \neq 0$ 
	we have
	\beqa
	&&E(z; p) = S^{(3)} ( z/\sqrt{p_*}), \qquad F(z; p) = S^{(4)} ( \sqrt{p} z), \\
	\label{spectralshift}
	&& K^{+}(z; p) = \psi^{(34)}(\sqrt{p_*} z), \qquad K^{-}(z; p) = \psi^{(34)}(z/ \sqrt{p}).
	\eeqa
	Note that after the quasi-Hopf twist $K^{\pm}(z)$ are related by the shift of the spectral parameter.
	Note also that \eqref{spectralshift} is valid only after the quasi-Hopf twist, since it involves $p \neq 0$.


\end{document}